\documentclass[english,twocolumn,pra,eqsecnum, showpacs]{revtex4}
\usepackage[T1]{fontenc}
\usepackage[latin1]{inputenc}
\usepackage{amsmath}
\usepackage{graphicx}
\usepackage{amssymb}

\makeatletter

\makeatletter

\makeatletter
\usepackage{epsf}

\makeatother

\makeatother

\usepackage{babel}
\makeatother

\begin{document}

\title{Fulde-Ferrell-Larkin-Ovchinnikov states in one-dimensional spin-polarized
ultracold atomic Fermi gases}

\author{Xia-Ji Liu$^{1,2}$, Hui Hu$^{1,2}$, and Peter D. Drummond${^{1}}$}

\affiliation{$^{1}$\ ARC Centre of Excellence for Quantum-Atom Optics, School
of Physical Sciences, University of Queensland, Brisbane, Queensland
4072, Australia \\
 $^{2}$\ Department of Physics, Renmin University of China, Beijing
100872, China}

\date{\today{}}

\begin{abstract}
We present a systematic study of quantum phases in a one-dimensional
spin-polarized Fermi gas. Three comparative theoretical methods are
used to explore the phase diagram at zero temperature: the mean-field
theory with either an order parameter in a single-plane-wave form
or a self-consistently determined order parameter using the Bogoliubov-de
Gennes equations, as well as the exact Bethe ansatz method. We find
that a spatially inhomogeneous Fulde-Ferrell-Larkin-Ovchinnikov phase,
which lies between the fully paired BCS state and the fully polarized
normal state, dominates most of the phase diagram of a uniform gas.
The phase transition from the BCS state to the Fulde-Ferrell-Larkin-Ovchinnikov
phase is of second order, and therefore there are no phase separation
states in one-dimensional homogeneous polarized gases. This is in
sharp contrast to the three-dimensional situation, where a phase separation
regime is predicted to occupy a very large space in the phase diagram.
We conjecture that the prediction of the dominance of the phase separation
phases in three dimension could be an artifact of the non-self-consistent
mean-field approximation, which is heavily used in the study of three-dimensional
polarized Fermi gases. We consider also the effect of a harmonic trapping
potential on the phase diagram, and find that in this case the trap
generally leads to phase separation, in accord with the experimental
observations for a trapped gas in three dimension. We finally investigate
the local fermionic density of states of the Fulde-Ferrell-Larkin-Ovchinnikov
ansatz. A two-energy-gap structure is shown up, which could be used
as an experimental probe of the Fulde-Ferrell-Larkin-Ovchinnikov states.
\end{abstract}

\pacs{03.75.Ss, 05.30.Fk, 71.10.Pm, 74.20.Fg}

\maketitle

\section{Introduction}

Since the successful demonstration of a magnetic Feshbach resonance\citep{FR}
and the creation of optical lattices\citep{lattice}, ultracold atomic
Fermi gases have become a topic of great current interest\citep{huinp}.
Thanks to these key tools, the inter-atomic interactions and even
the dimensionality of ultracold atomic Fermi gases can be easily tuned,
which makes them ideal candidates to simulate novel quantum many-particle
systems. Therefore, an intriguing opportunity is opened for studying
some long-standing problems, such as the crossover from Bardeen-Cooper-Schrieffer
(BCS) superfluidity to Bose-Einstein condensate (BEC) \citep{leggett,nsr,randeria,griffin,hui04,hld},
and models of high temperature superconductivity. These remarkable
prospects have attracted attention from many researchers, ranging
from condensed matter physics to atomic molecular and optical physics,
and even particle and astro physics. Experimentally, superfluidity
of an ultracold Fermi gas at the strongly interacting BCS-BEC crossover
has been strikingly demonstrated\citep{jila,mit04,duke04,ins04a,ins04b,ens,duke05a,mit05,duke05b,mit06,duke07,randy}.
This is a landmark achievement in the history of physics.

Recent experiments have now generated ultracold atomic Fermi gases
with finite spin polarization\citep{mit06a,mit06b,mit06c,mit07,rice06a,rice06b}.
That is, the two spin components have unequal populations. However,
the physical understanding of the ground state of a polarized atomic
gas remains an open question. The standard BCS model - though not
quantitatively accurate for strong interactions - is still qualitatively
correct when there is no spin polarization. This simply involves Cooper
pairing between spin up and spin down atoms with opposite momenta
at the same Fermi surface. A polarized Fermi gas cannot be explained
within standard BCS theory because the Fermi surfaces of the two spin
components are mismatched. Non-standard forms of pairing must exist
to support superfluidity in this polarized environment.

The study of polarized Fermi gases can be traced back to the middle
of the twentieth century, soon after the seminal BCS theory paper.
Similar theoretical proposals were independently given by Fulde and
Ferrell \citep{ff}, and Larkin and Ovchinnikov \citep{lo} (FFLO).
These authors suggested that Cooper pairs may acquire a finite center-of-mass
momentum \citep{rmp}. In such an ansatz, the two mismatched Fermi
surfaces can overlap, thereby supporting a spatially inhomogeneous
superfluidity. The search for the existence of the predicted FFLO
state has lasted for more than four decades. Only very recently has
there been indirect experimental evidence for observing such states
in the heavy fermion superconductor CeCoIn$_{5}$ \citep{cecoin5}.
Due to the shrinkage of the available phase space for pairing, the
FFLO state is now thought to be very fragile in three dimensions.
Alternative pairing scenarios include: Sarma superfluidity \citep{sarma,yip03,yip06},
a deformed Fermi surface \citep{dfs02,dfs05,dfs06}, and breached
pairing \citep{bp}. However, at zero temperature these phases may
suffer from an instability towards phase separation. As a result,
a phase separation regime consisting of a conventional BCS superfluid
and a normal fluid may be favored in three dimensions \citep{bedaque}.

The above theoretical issues were not completely resolved in current
measurements on polarized $^{6}$Li gases near a broad Feshbach resonance,
carried out at MIT \citep{mit06a,mit06b,mit06c,mit07} and Rice university
\citep{rice06a,rice06b}. Though a clear quantum phase transition
from a superfluid to normal state was observed \citep{mit06a}, the
nature and the order of the transition could not be determined due
to the finite experimental resolution. The presence of a harmonic
trap in these experiments caused additional difficulties in interpreting
the experimental results. A number of theoretical papers have sought
to explain these experiments on polarized atomic Fermi gases \citep{son,mannarelli,yang1,yang2,srprl,sraop,chevy1,chevy2,hui06,xiaji06,hui07,xiaji07,yip1,yip2,yip3,parish,lobo,torma1,torma2,torma3,bulgac1,bulgac2,levin1,levin2,carlson,lianyihe,caldas1,caldas2,ho,gu,iskin,duan1,duan2,duan3,duan4,silva1,silva2,stoof1,stoof2,stoof3,ldabec,martikainen,castorina,machida1,machida2}.
>From these analyses, the issues that require timely clarification
may be summarized as follows:

\textbf{(A)} {\em Structure and detection of FFLO states}. Despite
a long history, the precise structure of the FFLO states remains elusive
\citep{rmp}. Current investigations of FFLO states rely mostly on
the use of a single-plane-wave form for the pairing order parameter
$\Delta({\bf x})$, where $\Delta({\bf x})=\Delta_{0}\exp[i{\bf q\cdot x}]$,
as initially proposed by Fulde and Ferrell \citep{ff} (FF). Here
${\bf q}$ is the center-of-mass momentum of the Cooper pairs, and
the ansatz implies that the magnitude of the order parameter and density
is constant in space \citep{srprl,hui06,lianyihe}. The resulting
window for the FFLO state in parameter space turns out to be very
narrow \citep{srprl}. Can we expect a larger parameter range after
an optimization of the FFLO proposal? Indeed, by improving the form
of the order parameter to the Larkin and Ovchinnikov (LO) type, $\Delta({\bf x})\propto\cos[{\bf q\cdot x}]$,
Yoshida and Yip have found recently that the FFLO state became more
stable \citep{yip3}. On the other hand, so far there is no conclusive
evidence for the experimental observations of FFLO states \citep{rmp}.

\textbf{(B)} {\em Intrinsic reason for phase separation}. The narrow
window of the FFLO state may require phase separation to fill the
gap between BCS and FFLO phases in the phase diagram \citep{bedaque}.
Experimentally, a shell structure in the density profile of polarized
Fermi gases was observed \citep{mit06b,rice06a}, suggesting an interior
core of a BCS superfluid state with an outer shell of the normal component.
Phase separation in trapped systems, however, cannot be used as a
definitive support of the existence of phase separation in a homogeneous
gas, since the trap favors separation.

\textbf{(C)} {\em Quantitative approach for polarized Fermi gases
at the BCS-BEC crossover}. A more serious problem is the validity
of the mean-field approach. The experiments were done in the strongly
interacting BCS-BEC crossover regime, where for the quantitative purpose
strong pair fluctuations must be taken into account \citep{nsr,randeria,griffin,hld}.
Because of the lack of reliable knowledge of the superfluid phase,
these pair fluctuations are usually only considered above the superfluid
transition temperature \citep{xiaji06,parish}. For the same reason,
numerical quantum Monte Carlo simulations have been restricted to
the normal state \citep{lobo,carlson} and hence cannot provide useful
information for the superfluid state.

To gain a qualitative insight into these crucial points, in a recent
Letter \citep{hldprl1d}, we have considered a polarized Fermi gas
in one dimension (1D) at zero temperature. In this case the model
in free space is exactly soluble via a Bethe ansatz solution \citep{gaudin,takahashi,krivnov,guan1,guan2,xiaji1d,orso}.
We have established the 1D phase diagram of the polarized gas, both
in the uniform situation and in the experimentally important trapped
environment. Complemented by a mean-field Bogoliubov-de Gennes (BdG)
calculation, we have shown that a phase similar to the FFLO-type polarized
superfluid is the most widespread in the phase diagram. Using a local
density approximation to account for the harmonic trapping potential,
we have found that the trap generally leads to phase separation, with
at least one FFLO-type phase present at the trap center.

In this paper, we discuss these results in greater detail, and compare
them to other approximations. We particularly focus on the self-consistent
BdG method, which we previously treated briefly\citep{hldprl1d}.
To address the issue of the different possible FFLO structures, we
present a simplified mean-field calculation with a single-plane-wave
assumption for the order parameter, and compare it with the self-consistent
BdG results. These systematic investigations give rise to a comprehensive
quantitative understanding of the 1D polarized Fermi gas. We note
that a qualitative picture was also obtained in earlier some works,
which were based on a non-perturbative bosonization analysis \citep{yang1d}
or a mean-field approximation with an additional assumption on the
single-particle energy spectrum \citep{mf1d,buzdin}. However, the
resulting phase diagram was not conclusive, and the nature of the
transition from BCS to FFLO states was under debate \citep{yang1d}.

Strictly speaking, any mean-field approach is only valid in the weak
coupling limit. As the interaction strength increases, the pair fluctuations
become increasingly important, and therefore have to be taken into
account. This is particularly noticeable in 1D, where true long-range
order is completely destroyed by fluctuations in a homogeneous system
in the thermodynamic limit \citep{yang1d}, according to the well-known
Hohenberg-Mermin-Wagner theorem. To avoid this technical difficulty,
we therefore understand that the polarized gas under study is confined
either in a box with a finite length $L$ or in a harmonic trap (following
the experiments), although sometimes we would like to extend the length
$L$ to infinity.

The key results of the present work are that the structure of the
1D FFLO state is clarified. The transition from the BCS state to the
FFLO state is shown to be smooth, in marked contrast to the prediction
of a first order transition in 3D \citep{srprl}. Therefore, a 1D
phase separation is excluded in the phase diagram of the uniform system.
The phase separation in traps found in our previous Letter is indeed
simply an artifact of the parabolic trap, as we anticipated. It is
possible that similar effects are responsible for the phase separation
observations in the Rice experiment \citep{rice06a}, which uses a
high aspect ratio, elongated 3D trap.

It should be emphasized that as well as being an instructive theoretical
test bed for the ground state problem for a 3D gas, a 1D polarized
Fermi gas in a trap can be realized exactly using two-dimensional
optical lattices \citep{esslinger1,esslinger2}. In these experiments
the radial motion of atoms is frozen to zero-point oscillations due
to a tight transverse confinement, while the axial motion is weakly
confined. Thus, one can realize a low-dimensional quantum many-body
system, and experimentally check the many-body predictions directly.
This has also been recently carried out for a 1D Bose gas \citep{esslinger1,peter}.

The paper is organized as follows. In the following section, we outline
the theoretical model for a 1D spin-polarized Fermi gas. In Sec. III,
we characterize the uniform phase diagram by using a simplified mean-field
approach with a single-plane-wave like order parameter, \textit{i.e.},
the so-called FF solution for the FFLO state. This provides us with
an approximate picture of the ground state of a 1D polarized gas.
An improved self-consistent BdG calculation is then given in Sec.
IV, without any assumption for the order parameter. The underlying
structure of the FFLO states at all spin polarizations is then analyzed.
The comparison between these two different mean-field approaches shows
that the simple FF ansatz fails to capture the correct physics around
the BCS-FFLO transition point. It therefore predicts the wrong type
of transition. We conclude that in a 3D polarized gas case, the FF
ansatz could lead to the same incorrect conclusion. In Sec. V the
validity of these 1D mean-field analyses in the weak-coupling or intermediate-coupling
regime is checked using exact Bethe ansatz solutions. A quantitative
phase diagram of a homogeneous gas is obtained by gathering all the
information from these three methods.

In Secs. VI and VII we study the trapped case, using either the self-consistent
BdG equations or the exact solution within the local density approximation.
We again find a good agreement between these two results for weak
and moderate couplings. The phase diagram of the trapped gas is thereby
determined. We also calculate the local fermionic density of states
of the FFLO states. A two-energy-gap structure is predicted, which
is potentially useful for the experimental detection of FFLO states.
Finally, Sec. VIII is devoted to the conclusions and some final remarks.

\section{Models}

Consider a polarized Fermi gas with a broad Feshbach resonance in
a highly elongated trap formed using a two dimensional optical lattice
\citep{esslinger1}. By suitably tuning the lattice depth, the anisotropy
aspect ratio $\lambda=\omega_{z}/\omega_{\rho}$ of two harmonic frequencies
can be extremely small. As long as the Fermi energy associated with
the longitudinal motion of the atoms is much smaller than the energy
level separation along the transverse direction, \textit{i.e.}, $k_{B}T\ll\hbar\omega_{\rho}$
and $N\hbar\omega_{z}\ll\hbar\omega_{\rho}$, where $N$ is the total
number of atoms, the transverse motion will be essentially frozen
out. One ends up with a quasi-one dimensional system. The effective
Hamiltonian of the 1D polarized attractive Fermi gas then may be described
by a single channel model \citep{karen,randy,diener,xiaji},

\begin{eqnarray}
H & = & \mathop{\textstyle \sum}\limits _{\sigma}\int dx{\textstyle \Psi_{\sigma}^{+}\left(x\right)\left[-\frac{\hbar^{2}\nabla^{2}}{2m}+V_{trap}\left(x\right)-\mu_{\sigma}\right]\Psi_{\sigma}\left(x\right)}\nonumber \\
 & + & g_{1D}{\textstyle \int dx\Psi_{\uparrow}^{+}\left(x\right)\Psi_{\downarrow}^{+}\left(x\right)\Psi_{\downarrow}\left(x\right)\Psi_{\uparrow}\left(x\right),\label{model}}\end{eqnarray}
 where the pseudospins $\sigma=\uparrow,\downarrow$ denote the two
hyperfine states, and $\Psi_{\sigma}\left(x\right)$ is the Fermi
field operator that annihilates an atom at position $x$ in the spin
$\sigma$ state. The number of atoms in each spin component is $N_{\sigma}$
and the total number of atoms is $N=N_{\uparrow}+N_{\downarrow}$.
Two different chemical potentials, $\mu_{\uparrow,\downarrow}=\mu\pm\delta\mu$,
are introduced to take into account the population imbalance $\delta N=N_{\uparrow}-N_{\downarrow}>0$.
The potential $V_{trap}\left(x\right)=m\omega^{2}x^{2}/2$ defines
a harmonic trap with an oscillation frequency $\omega=\omega_{z}$
in the axial direction. In such a quasi-one dimensional geometry,
it is shown by Bergeman \textit{et al.} \citep{bergeman} that the
scattering properties of the atoms can be well described using a contact
potential $g_{1D}\delta(x)$, where the 1D effective coupling constant
$g_{1D}<0$ may be expressed through the 3D scattering length $a_{3D}$,

\begin{equation}
g_{1D}=\frac{2\hbar^{2}a_{3D}}{ma_{\rho}^{2}}\frac{1}{\left(1-Aa_{3D}/a_{\rho}\right)}.\label{eq:G1D}\end{equation}
 Here $a_{\rho}=\sqrt{\hbar/(m\omega_{\rho})}$ is the characteristic
oscillator length in the transverse axis, and the constant $A=-\zeta(1/2)/\sqrt{2}\simeq1.0326$
is responsible for the confinement induced Feshbach resonance \citep{bergeman,astrakharchik,footnote},
which changes the scattering properties dramatically when the 3D scattering
length is comparable to the transverse oscillator length. It is also
convenient to express $g_{1D}$ in terms of an effective 1D scattering
length, $g_{1D}=-2\hbar^{2}/\left(ma_{1D}\right)$, where \begin{equation}
a_{1D}=-\frac{a_{\rho}^{2}}{a_{3D}}\left(1-A\frac{a_{3D}}{a_{\rho}}\right)>0.\end{equation}
 Note that in the definition of the 1D scattering length, the sign
convention is opposite to the 3D case.

In this paper, we will assume a negative 3D scattering length. In
other words, the 1D attractive polarized Fermi gas would be obtained
experimentally from a 3D polarized gas on the BCS side of the Feshbach
resonance magnetic field \citep{tokatly,zwerger}.

In the absence of the harmonic trap, we measure the interactions by
a dimensionless parameter $\gamma$, which is the ratio of the interaction
energy density $e_{int}$ to the kinetic energy density $e_{kin}$
\citep{lieb}. In the weak coupling limit, $e_{int}\sim g_{1D}n$
and $e_{kin}\sim\hbar^{2}k^{2}/(2m)\sim\hbar^{2}n^{2}/m$, where $n$
is the total linear density. Therefore, one finds \begin{equation}
\gamma=-\frac{mg_{1D}}{\hbar^{2}n}=\frac{2}{na_{1D}}\end{equation}
 Thus, $\gamma\ll1$ corresponds to the weakly interacting limit,
while the strong coupling regime is realized when $\gamma\gg1$.

In the case of a trap, we may characterize the interactions using
the dimensionless parameter at the trap center $\gamma_{0}=\gamma(x=0)$.
For an ideal two-component Fermi gas with equal spin populations,
the total linear density is \begin{equation}
n_{ideal}\left(x\right)=n_{0}\left(1-\frac{x^{2}}{x_{TF}^{2}}\right)^{1/2},\end{equation}
 in the Thomas-Fermi (TF) approximation, where \begin{eqnarray}
n_{0} & = & \frac{2N^{1/2}}{\pi a_{ho}},\\
x_{TF} & = & N^{1/2}a_{ho},\end{eqnarray}
 are respectively the center linear density and the TF radius. Here
$a_{ho}=\sqrt{\hbar/(m\omega_{z})}$ is the characteristic oscillator
length in the axial direction. We thus estimate \begin{equation}
\gamma_{0}=\frac{\pi}{N^{1/2}}\left(\frac{a_{ho}}{a_{1D}}\right).\end{equation}

In our previous Letter \citep{hldprl1d}, we have defined a dimensionless
quantity $\kappa=Na_{1D}^{2}/a_{ho}^{2}$ to describe the interactions.
These are related via $\gamma_{0}=\pi/(\sqrt{\kappa})$.

Finally, we use a capital $P=(N_{\uparrow}-N_{\downarrow})/N$ to
label the total spin polarization, and $p=(n_{\uparrow}-n_{\downarrow})/n$
to denote the local (or uniform) spin polarization.

%
\begin{figure}
\begin{centering}
\includegraphics[clip,width=0.45\textwidth]{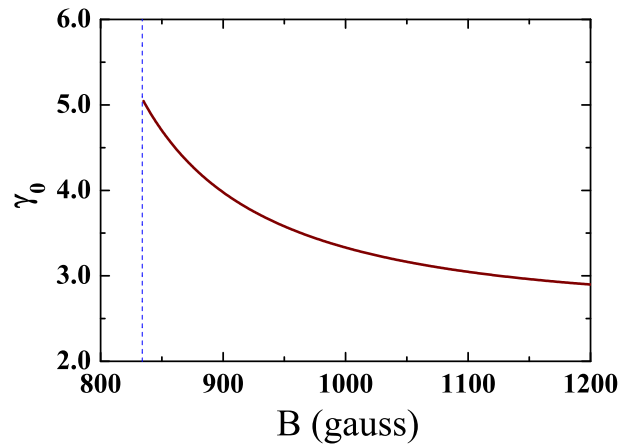}
\par\end{centering}

\caption{(Color online) Dimensionless coupling constant at the trap center
as a function of the magnetic field. This plot is designed specifically
to represent a polarized gas of $^{6}$Li atoms in a two-dimensional
optical lattice, assuming the same conditions as in the MIT experiment
\citep{mit06}. In detail, we take the total number of atoms as $\sim10^{5}$,
and therefore in each tube the number of fermions is about $N\sim100$.
The periodicity of the lattice is $d=532$nm, yielding a transverse
scale $a_{\rho}\simeq120$nm. The axial confinement frequency $\omega\sim2\pi\times400$Hz,
giving rise to an axial oscillator length $a_{ho}\simeq2\mu m$. The
3D scattering length is related to the magnetic field via $a_{3d}=-1405a_{0}[1+300/(B-834)][1+0.0004(B-834)]$,
where the magnetic field $B$ is measured in Gauss and $a_{0}=0.0529$nm
is the Bohr radius. The dashed line in the figure shows the Feshbach
resonance field.}

\label{fig1}
\end{figure}


To make the experimental relevance, we estimate the dimensionless
interaction parameters for the on-going experiments on one-dimensional
polarized Fermi gases. A gas of $^{6}$Li atoms in a three-dimensional
optical lattice has been successfully produced by the MIT group \citep{mit06}.
Thus, we consider the case of $^{6}$Li gas loaded into a two-dimensional
optical lattice with the same parameters. Typically, in each one-dimensional
tube the number of $^{6}$Li atom is about $N\sim100$. The transverse
oscillator length $a_{\rho}$ is related to the periodicity of the
lattice $d$ via $a_{\rho}=d/(\pi s^{1/4})$ \citep{zwerger03}, where
$s$ is the ratio of the lattice depth to the recoil energy. Taking
$s=4$, the experimental value of $d=532$nm then yields $a_{\rho}\simeq120$nm.
An axial confinement of $\omega\sim2\pi\times400$Hz gives rise to
an axial oscillator length $a_{ho}=\sqrt{\hbar/(m\omega)}\simeq2\mu m$.
Further, the three-dimension scattering length of $^{6}$Li gas at
the broad resonance is given by \citep{a3dB}, $a_{3d}=-1405a_{0}[1+300/(B-834)][1+0.0004(B-834)]$,
where the magnetic field $B$ is measured in Gauss and $a_{0}=0.0529$nm
is the Bohr radius. We then use the relation, \begin{equation}
\gamma_{0}=-\frac{\pi}{N^{1/2}}\frac{a_{ho}a_{3D}}{a_{\rho}^{2}}\frac{1}{\left(1-Aa_{3D}/a_{\rho}\right)},\end{equation}
 to estimate the dimensionless coupling constant at the trap center.

Fig. 1 gives the resulting $\gamma_{0}$ as a function of the magnetic
field $B$. We find that $\gamma_{0}\sim O(1)$ above the Feshbach
resonance. Throughout this work we shall take a coupling constant
of $\gamma=1.6$. We note that there is already some indirect evidence
for superfluidity of a Fermi gas in a three-dimensional optical lattice
\citep{mit06}, at the magnetic field considered. On switching to
a two-dimensional optical lattice, the temperature in the experiments
may still be low enough to generate the various one-dimensional superfluid
phases at zero temperature.

Throughout the paper we shall mainly study two different cases, either
with a fixed total number of particles and a fixed chemical potential
difference, or with given numbers of both spin-up and spin-down particles.
The system with two fixed chemical potentials may be considered as
well. These three situations require the use of different canonical
ensembles in thermodynamics. In the first two cases, we minimize the
free energies of the system, $F_{\delta\mu}(T,V,n,\delta\mu)$ and
$F_{\delta n}(T,V,n,\delta n)$, respectively. While in the latter
case, we minimize instead the thermodynamic potential $\Omega(T,V,\mu,\delta\mu)$.

\section{Single plane wave approximation in a homogeneous gas}

We first consider a mean-field description with a single plane-wave
FF type order parameter, to give the simplest qualitative picture
of a homogeneous polarized Fermi gas \citep{hui06}. At this point,
we write the Hamiltonian (\ref{model}) in momentum space using a
Fourier decomposition of the Fermi field operators. This results in:
\begin{eqnarray}
{\cal H}_{\hom} & = & \sum_{k\sigma}\left(\epsilon_{k}-\mu_{\sigma}\right)c_{k\sigma}^{+}c_{k\sigma}\nonumber \\
 &  & +g_{1D}\sum_{pkk^{\prime}}c_{p/2+k\uparrow}^{+}c_{p/2-k\downarrow}^{+}c_{p/2-k^{\prime}\downarrow}c_{p/2+k^{\prime}\uparrow,}\end{eqnarray}
 where $\epsilon_{k}=\hbar^{2}k^{2}/2m$ is the kinetic energy. The
single-plane-wave mean-field approximation amounts to decoupling the
interaction term using an order parameter $\Delta=-g_{1D}\sum_{k}\left\langle c_{q/2-k\downarrow}c_{q/2+k\uparrow}\right\rangle $
for the Cooper pairs, where we assume that the pairing occurs between
a spin up atom with a momentum $q/2+k$ and a spin down atom with
a momentum $q/2-k$. As a result, the pairs possess a specific nonzero
center-of-mass momentum $q$, whose value, together with the value
of $\Delta$, are to be determined. It is easy to see that after a
Fourier transformation, the order parameter in real space acquires
a single-plane-wave form, \textit{i.e.}, $\Delta(x)=\Delta\exp[iqx]$.
Therefore, within this approximation, we have a mean-field Hamiltonian,
\begin{eqnarray}
{\cal H}_{\hom}^{MF} & = & {\cal -}\frac{\Delta^{2}}{g_{1D}}-g_{1D}n_{\uparrow}n_{\downarrow}+\sum_{k\sigma}\left(\epsilon_{k}-\tilde{\mu}_{\sigma}\right)c_{k\sigma}^{+}c_{k\sigma}\nonumber \\
 &  & -\Delta\sum_{k}\left(c_{q/2-k\downarrow}c_{q/2+k\uparrow}+h.c.\right).\end{eqnarray}
 Here, as a consequence of the constant linear density, Hartree terms
like $g_{1D}n_{-\sigma}c_{k\sigma}^{+}c_{k\sigma}$ merely introduce
an overall shift for the chemical potentials. We indicate this by
introducing the notation $\tilde{\mu}_{\sigma}=\mu_{\sigma}-g_{1D}n_{-\sigma}$
for the shifted chemical potentials.

To solve the mean-field Hamiltonian, it is convenient to use a Nambu
spinor creation operator $\psi_{k}^{+}=(c_{q/2+k\uparrow}^{+},c_{q/2-k\downarrow})$.
The Hamiltonian may then be rewritten in a compact bilinear form,
\begin{eqnarray}
{\cal H}_{\hom}^{MF} & = & \mathop{\textstyle \sum}_{k}\psi_{k}^{+}\left[\left(\epsilon_{k}^{+}-\tilde{\mu}\right){\bf \sigma}_{z}-\Delta{\bf \sigma}_{x}+\left(\epsilon_{k}^{-}-\delta\tilde{\mu}\right)\right]\psi_{k}\nonumber \\
 &  & {\cal -}\frac{\Delta^{2}}{g_{1D}}-g_{1D}n_{\uparrow}n_{\downarrow}+\mathop{\textstyle \sum}_{k}\left(\epsilon_{k}-\tilde{\mu}+\delta\tilde{\mu}\right),\label{mfhami}\end{eqnarray}
 where $\epsilon_{k}^{\pm}=(\epsilon_{q/2+k}\pm\epsilon_{q/2-k})/2$,
and ${\bf \sigma}_{x}$ and ${\bf \sigma}_{z}$ are the Pauli matrices.
For convenience, we have defined, \begin{eqnarray}
\tilde{\mu} & = & \mu-\frac{g_{1D}n}{2},\\
\delta\tilde{\mu} & = & \delta\mu+\frac{g_{1D}\delta n}{2}.\end{eqnarray}
 The bilinear Hamiltonian can be easily diagonalized by working out
the eigenvalues $E_{k}^{\pm}$ and eigenstates $\Phi_{k}^{\pm}$ of
the two by two matrix $[(\epsilon_{k}^{+}-\tilde{\mu}){\bf \sigma}_{z}-\Delta{\bf \sigma}_{x}+(\epsilon_{k}^{-}-\delta\tilde{\mu})]$.
Explicitly, we find that \begin{equation}
E_{k}^{\pm}=\epsilon_{k}^{-}-\delta\tilde{\mu}\pm E_{k},\label{spw-energies}\end{equation}
 and \begin{equation}
\Phi_{k}^{+}=\left(\begin{array}{c}
u_{k}\\
v_{k}\end{array}\right),\quad\Phi_{k}^{-}=\left(\begin{array}{c}
-v_{k}^{*}\\
u_{k}^{*}\end{array}\right),\label{spw-wvfs}\end{equation}
 where $E_{k}=[(\epsilon_{k}^{+}-\tilde{\mu})^{2}+\Delta^{2}]^{1/2}$
and \begin{eqnarray}
u_{k}^{2} & = & \frac{1}{2}\left[1+\frac{\epsilon_{k}^{+}-\tilde{\mu}}{E_{k}}\right],\\
v_{k}^{2} & = & \frac{1}{2}\left[1-\frac{\epsilon_{k}^{+}-\tilde{\mu}}{E_{k}}\right],\\
u_{k}v_{k} & = & -\frac{\Delta}{2E_{k}}.\end{eqnarray}
 From the eigenstates $\Phi_{k}^{\pm}$, it is natural to define Bogoliubov
quasiparticle operators, which are given by: \begin{equation}
\left(\begin{array}{c}
\alpha_{k\uparrow}\\
\alpha_{-k\downarrow}^{+}\end{array}\right)=\left(\begin{array}{cc}
u_{k}, & v_{k}^{*}\\
-v_{k}, & u_{k}^{*}\end{array}\right)\psi_{k}.\end{equation}
 The bilinear mean-field Hamiltonian then becomes \begin{eqnarray}
{\cal H}_{\hom}^{MF} & = & {\cal -}\frac{\Delta^{2}}{g_{1D}}-g_{1D}n_{\uparrow}n_{\downarrow}+\sum_{k}\left(\epsilon_{k}^{+}-\tilde{\mu}-E_{k}\right)\nonumber \\
 &  & +\sum_{k}\left[E_{k}+\epsilon_{k}^{-}-\delta\tilde{\mu}\right]\alpha_{k\uparrow}^{+}\alpha_{k\uparrow}\nonumber \\
 &  & +\sum_{k}\left[E_{k}-\epsilon_{k}^{-}+\delta\tilde{\mu}\right]\alpha_{k\downarrow}^{+}\alpha_{k\downarrow}.\end{eqnarray}

The thermodynamic potential is obtained by replacing $\alpha_{k\sigma}^{+}\alpha_{k\sigma}$
by its thermal statistical average values, \textit{i.e.}, the Fermi
distribution function $f(E_{k}^{\pm})=1/(\exp[\beta E_{k}^{\pm}]+1)$
with $\beta=1/(k_{B}T)$ as the inverse temperature. At zero temperature
where $\beta$ goes to infinity, the Fermi distribution function $f(x)$
reduces to a step function $\Theta\left(-x\right)$, \textit{i.e.},
$\Theta\left(x>0\right)=1$ and $\Theta\left(x<0\right)=0$, so the
resulting thermodynamic potential has the form: \begin{eqnarray}
\Omega & = & {\cal -}\frac{\Delta^{2}}{g_{1D}}-g_{1D}n_{\uparrow}n_{\downarrow}+\sum_{k}\left(\epsilon_{k}^{+}-\tilde{\mu}-E_{k}\right)\nonumber \\
 &  & +\sum_{k}\left[E_{k}+\epsilon_{k}^{-}-\delta\tilde{\mu}\right]\Theta\left(-E_{k}^{+}\right)\nonumber \\
 &  & +\sum_{k}\left[E_{k}-\epsilon_{k}^{-}+\delta\tilde{\mu}\right]\Theta\left(-E_{k}^{-}\right),\end{eqnarray}

The values of the order parameter $\Delta$ and of the pairing momentum
$q$ are determined by finding the stationary points in the $(\Delta,q)$
plane of the thermodynamic potential, \textit{i.e.}, $\partial\Omega/\partial\Delta=0$
and $\partial\Omega/\partial q=0$, with given chemical potential
difference $\delta\mu$, or the requirement of number conservation,
$\delta n=-\partial\Omega/\partial\delta\mu$. This gives us two distinct
procedures for defining the mean-field solution, analogous to the
grand-canonical (fixed chemical potential difference) and canonical
(fixed number difference) ensembles in thermodynamics.

Once these variational variables are obtained, we calculate straightforwardly
the total free energies $F_{\delta\mu}=\Omega+\mu n=\tilde{F}_{\delta\mu}+g_{1D}(n^{2}+\delta n^{2})/4$
or $F_{\delta n}=\Omega+\mu n+\delta\mu\delta n=\tilde{F}_{\delta n}+g_{1D}(n^{2}-\delta n^{2})/4$
of the gas, depending on whether the chemical potential difference
$\delta\mu$ or the number difference $\delta n=n_{\uparrow}-n_{\downarrow}$
is fixed, as indicated by the subscript. Note that at zero temperature
the value of the free energy $F_{\delta n}$ is equal to the total
ground state energy $E$. We have also defined two free energies $\tilde{F}_{\delta\mu}$
and $\tilde{F}_{\delta n}$ in the absence of the Hartree terms. In
the detailed calculations, for a uniform system we take respectively
the Fermi energy $\epsilon_{F}=\hbar^{2}k_{F}^{2}/(2m)$ and the Fermi
wave vector $k_{F}=\pi n/2$ (of a unpolarized ideal gas ) as the
units of the energy and of the momentum, by letting $\hbar=1$ and
$2m=1$.

\subsection{Qualitative phase diagrams}

%
\begin{figure}
\begin{centering}
\includegraphics[clip,width=0.45\textwidth]{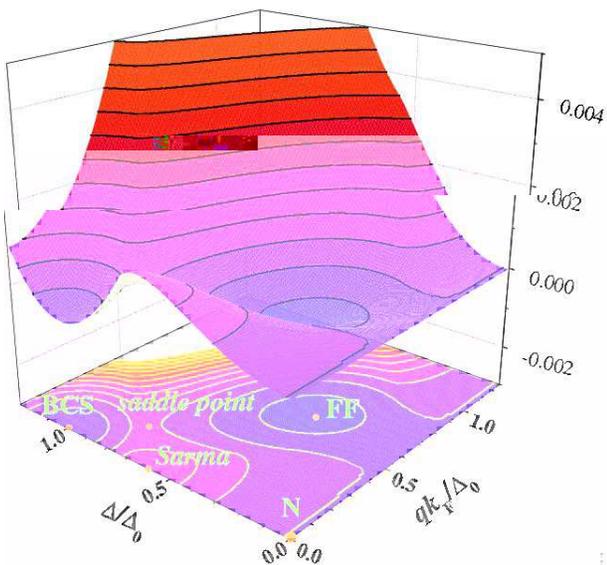}
\par\end{centering}

\caption{(Color online) Landscape of the zero-temperature thermodynamic potential
of a uniform gas at an interaction strength $\gamma=1.6$. Here, we
take a single-plane-wave approximation for the order parameter, and
normalize it using the full gap of an unpolarized Fermi gas, $\Delta_{0}=0.34658\epsilon_{F}$,
where $\epsilon_{F}$ is the Fermi energy. The chemical potential
is fixed at $\tilde{\mu}=1.04594\epsilon_{F}$. The competing ground
states are (i) a normal Fermi gas with $\Delta=0$, (ii) a fully paired
BCS superfluid with $\Delta=\Delta_{0}$, $q=0$, and $\delta n=0$,
(iii) a finite momentum paired FF superfluid with $\Delta<\Delta_{0}$,
$q\neq0$, and $\delta n\neq0$, (iv) a breached pairing or Sarma
superfluid with $\Delta<\Delta_{0}$, $q=0$, and $\delta n\neq0$,
and (v) a saddle point phase intervening between the local BCS and
FF minima. We note that the last two phases are unstable with respect
to phase separation.}

\label{fig2}
\end{figure}


Generally, there are several possible stationary solutions in the
landscape of the thermodynamic potential. On the weak coupling side
we find only three stable competing ground states, corresponding to
local minima of the landscape. As shown in Fig. 2 for a coupling constant
$\gamma=1.6$, these are the unpolarized (BCS), partially polarized
(FF), and a fully polarized or normal (N) phases. The other two states,
denoted as {}``Sarma'' and {}``saddle point'' phases, are unstable
with respect to phase separation \citep{hui06}. Note that in the
figure, the order parameter $\Delta$ and the center-of-mass momentum
$q$ are measured in units of the full gap of an unpolarized gas,
$\Delta_{0}\simeq0.34658\epsilon_{F}$. We have fixed the chemical
potential at its unpolarized value, $\tilde{\mu}\simeq1.04594\epsilon_{F}$,
and have taken the chemical potential difference to be $\delta\tilde{\mu}=0.75\Delta_{0}$.

%
\begin{figure}
\begin{centering}
\includegraphics[clip,width=0.45\textwidth]{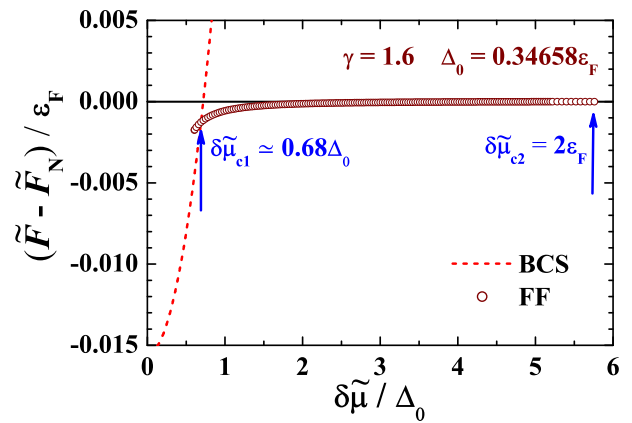}
\par\end{centering}

\caption{(Color online) Comparison of the free energies of $\tilde{F}_{\delta\mu}$
available mean-field solutions at a coupling constant $\gamma=1.6$
and at zero temperature, with the free energy of the normal gas $\tilde{F}_{N}$
being subtracted. With increasing the chemical potential difference,
the gas turns from a BCS superfluid to a FF superfluid at $\delta\tilde{\mu}\simeq0.68\Delta_{0}$,
and finally becomes a normal gas above $\delta\tilde{\mu}=2\epsilon_{F}$.}

\label{fig3}
\end{figure}


%
\begin{figure}
\begin{centering}
\includegraphics[clip,width=0.45\textwidth]{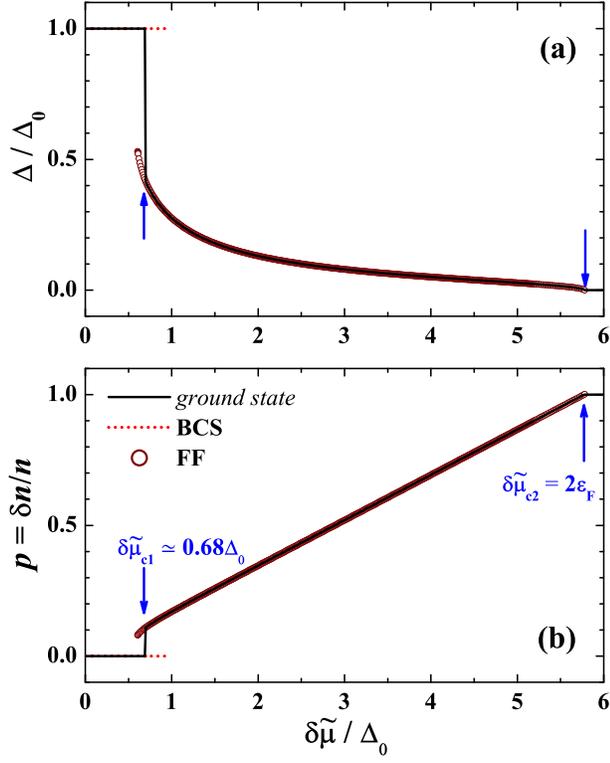}
\par\end{centering}

\caption{(Color online) Evolution of the mean-field (FF) order parameter and
of the spin polarization, with increasing chemical potential difference.
The arrows point to the phase transition positions. The parameters
are the same as in Fig. 3.}

\label{fig4}
\end{figure}


For an interaction strength $\gamma=1.6$, the evolution of the ground
states with increasing chemical potential difference is given in Fig.
3. Here we search for the ground state by minimizing the free energy
$F_{\delta\mu}$. As $\delta\tilde{\mu}$ increases from zero, the
free energy of the BCS state is initially lowest, but rises very rapidly.
It intersects with that of the FF state at about $\delta\tilde{\mu}=0.68\Delta_{0}$.
A first order quantum phase transition then occurs in mean-field theory,
since the first order derivative of free energies at the intersection
point is discontinuous. The apparent hysteresis (presence of the FF
state before the transition point) is also the mark of a first order
phase transition. After that, the free energy increases slowly towards
the normal state value. Precisely at $\delta\tilde{\mu}=2\epsilon_{F}$,
the gas enters smoothly into a fully polarized normal state, where
the spin polarization $p=(n_{\uparrow}-n_{\downarrow})/(n_{\uparrow}+n_{\downarrow})$
is strictly equal to one. Hence, differing from the 3D situation,
a partially polarized normal phase is excluded in 1D. We present,
respectively, the value of the order parameter and the spin polarization
as a function of the chemical potential difference in Figs. 4a and
4b. The first order transition from BCS to FF states becomes much
apparent due to the jump of the order parameter and of the spin polarization.
We will show later, however, that this apparent first order transition
is simply an artifact of the single-plane-wave approximation for the
order parameter.

%
\begin{figure}
\begin{centering}
\includegraphics[clip,width=0.45\textwidth]{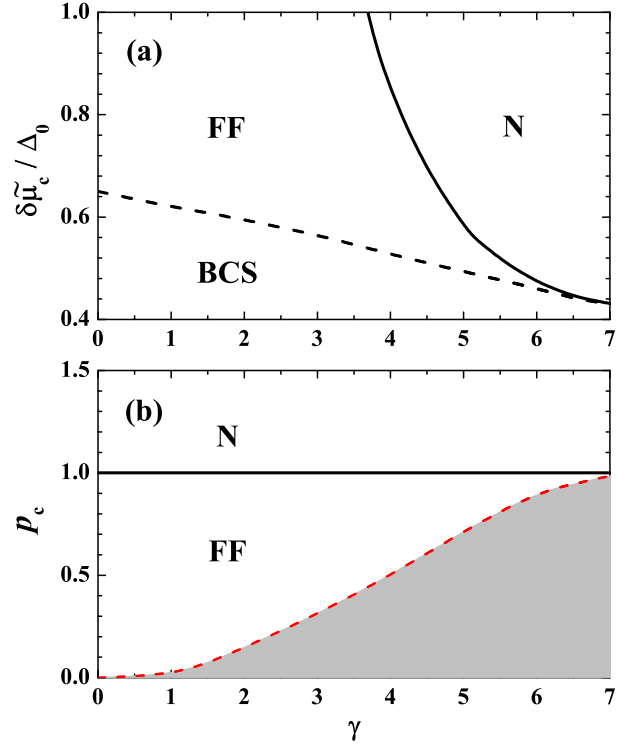}
\par\end{centering}

\caption{(Color online) (a) Phase diagram in the plane of the interaction strength
and the chemical potential difference. Within the single-plane-wave
assumption for the order parameter, the transition from a BCS superfluid
to a FF state is of first order (dashed line), while from a FF state
to the normal state it is continuous (solid line). (b) Interaction
strength vs polarization phase diagram. The shadow region is unknown,
and presumably is an artifact of the single-plane-wave approximation.}

\label{fig5}
\end{figure}


By changing the coupling constant, we can determine a phase diagram
in the plane of the interaction strength $\gamma$ and chemical potential
difference $\delta\tilde{\mu}$, as shown in Fig. 5a. The solid and
dashed lines separate the FF state from the normal and BCS phases
respectively, and converge to a single curve above $\gamma\simeq7$.
Converting the chemical potential difference to a number difference,
we obtain a phase diagram in the $\gamma-p$ plane in Fig. 5b. The
area under the dashed line has no correspondence in Fig. 5a and belongs
to the {}``saddle point'' solution, which is unstable towards phase
separation. This may be the precursor of a phase separation phase.
Overall, all the basic features found here are qualitatively similar
to that in 3D \citep{hui06}.

\subsection{Analytic results in limiting cases}

We discuss some analytic results that can be obtained in the weakly
interacting limit of $\gamma\rightarrow0$. The simplest one is the
unpolarized BCS state, for which the chemical potential $\tilde{\mu}$
is essentially the Fermi energy $\epsilon_{F}$. The stationary condition
$\partial\Omega/\partial\Delta=0$ then leads to a gap equation, \begin{equation}
\frac{1}{g_{1D}}+\sum_{k}\frac{1}{2\sqrt{\left(\epsilon_{k}-\epsilon_{F}\right)^{2}+\Delta_{0}^{2}}}=0.\end{equation}
 The integration can be worked out analytically for small $\Delta_{0}$.
One finds that \begin{equation}
\Delta_{0}\simeq8\epsilon_{F}\exp\left[-\frac{\pi^{2}}{2\gamma}\right],\end{equation}
 analogous to the standard 3D BCS result $\Delta_{0}^{3D}\simeq8\epsilon_{F}\exp[\pi/(2k_{F}a)-2]$.
For the FF state at a large chemical potential difference, the value
of the order parameter is even smaller. To a good approximation, we
find that \begin{eqnarray}
\tilde{\mu} & \simeq & \epsilon_{F}+\frac{\left(\delta\tilde{\mu}\right)^{2}}{4\epsilon_{F}},\\
qk_{F} & \simeq & \delta\tilde{\mu},\end{eqnarray}
 and hence: \begin{equation}
\Delta=8\epsilon_{F}\frac{\sqrt{\left(2\epsilon_{F}-\delta\tilde{\mu}\right)\left(2\epsilon_{F}+\delta\tilde{\mu}\right)}}{\delta\tilde{\mu}}\exp\left[-\frac{\pi^{2}}{\gamma}\right].\end{equation}
 From the prefactor, the order parameter $\Delta$ vanishes exactly
at $\delta\tilde{\mu}=2\epsilon_{F}$. At the same time $\tilde{\mu}=2\epsilon_{F}$,
indicating that the FF state changes smoothly into a fully polarized
normal state.

\subsection{Local fermionic density of states}

%
\begin{figure}
\begin{centering}
\includegraphics[clip,width=0.45\textwidth]{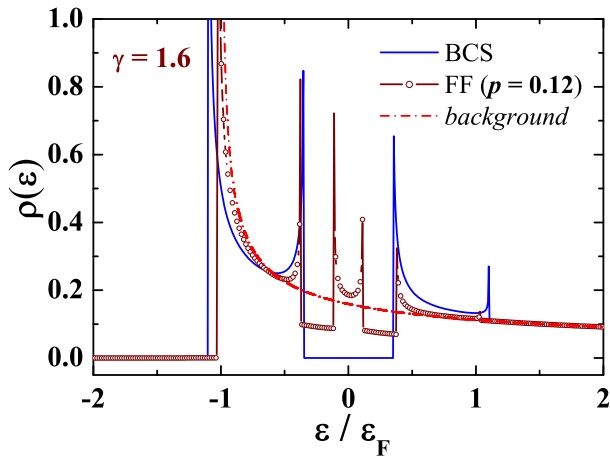}
\par\end{centering}

\caption{(Color online) Local fermionic density of states of a uniform polarized
Fermi gas, with a single-plane-wave form for the order parameter.
Note that there is a prominent two-energy-gap structure in the FF
state.}

\label{fig6}
\end{figure}


The Bogoliubov quasiparticle amplitudes $(u_{k},$ $v_{k})$ and energy
$E_{k}$ appear in the zero temperature spectrum of the single fermionic
excitations. We characterize the excitation spectrum using the local
fermionic density of states, $\rho_{\sigma}\left(\epsilon\right)$,
given by

\begin{eqnarray}
\rho_{\uparrow}(\epsilon) & = & \sum_{{\bf k}}u_{k}^{2}\delta\left(\epsilon-E_{k}^{+}\right)+\sum_{{\bf k}}v_{k}^{2}\delta\left(\epsilon-E_{k}^{-}\right),\\
\rho_{\downarrow}(\epsilon) & = & \sum_{{\bf k}}v_{k}^{2}\delta\left(\epsilon+E_{k}^{+}\right)+\sum_{{\bf k}}u_{k}^{2}\delta\left(\epsilon+E_{{\bf k}}^{-}\right).\end{eqnarray}
 For an ideal gas with equal populations, the density of states can
be calculated analytically, \begin{equation}
\rho_{\uparrow}^{bk}(\epsilon)=\rho_{\downarrow}^{bk}(\epsilon)=\frac{\sqrt{2m}}{2\pi\hbar}\frac{1}{\sqrt{\epsilon+\tilde{\mu}}}.\end{equation}
 which we have regarded as a background density of states. It has
a band edge (square root) singularity at $\epsilon=-\tilde{\mu}$.

We plot in Fig. 6 the local density of states for a one-dimensional
BCS superfluid, and the FF phase at $p=0.12$, as well as the background
density of states. In an FF state, the spin up and down density of
states are exactly the same, but are shifted downwards or upwards
respectively by an amount $\delta\tilde{\mu}$. For clarity, in the
figure we show only one branch, \textit{i.e.}, the spin up density
of states after an upwards shift. Compared to the BCS superfluid,
the local density of states of the FF phase exhibits an intriguing
two-energy-gap structure. The midgap state around $\epsilon=0$ is
a salient feature of the spatially modulated order parameter \citep{mf1d}.

\section{Self-consistent BdG in a homogeneous gas}

We now turn to a more realistic mean-field calculation without resorting
any approximation for the form of the order parameter. We consider
the BdG equations of the 1D polarized Fermi gas \citep{bdg,xiaji07},
starting from the Heisenberg equation of motion of the Hamiltonian
(\ref{model}) for $\Psi_{\uparrow}\left(x,t\right)$ and $\Psi_{\downarrow}\left(x,t\right)$
(without the trap potential):

\begin{eqnarray}
i\hbar\frac{\partial\Psi_{\uparrow}}{\partial t} & = & \left[-\frac{\hbar^{2}\nabla^{2}}{2m}-\mu_{\uparrow}\right]\Psi_{\uparrow}+g_{1D}\Psi_{\downarrow}^{+}\Psi_{\downarrow}\Psi_{\uparrow},\\
i\hbar\frac{\partial\Psi_{\downarrow}}{\partial t} & = & \left[-\frac{\hbar^{2}\nabla^{2}}{2m}-\mu_{\downarrow}\right]\Psi_{\downarrow}-g_{1D}\Psi_{\uparrow}^{+}\Psi_{\downarrow}\Psi_{\uparrow}.\end{eqnarray}
 Within the mean-field approximation, we replace the terms $g_{1D}\Psi_{\downarrow}^{+}\Psi_{\downarrow}\Psi_{\uparrow}$
and $g_{1D}\Psi_{\uparrow}^{+}\Psi_{\downarrow}\Psi_{\uparrow}$ by
their respective mean-field decoupling \begin{equation}
g_{1D}\Psi_{\downarrow}\Psi_{\uparrow}\Psi_{\downarrow}^{+}=-\Delta(x)\Psi_{\downarrow}^{+}+g_{1D}n_{\downarrow}(x)\Psi_{\uparrow},\end{equation}
 and \begin{equation}
g_{1D}\Psi_{\downarrow}\Psi_{\uparrow}\Psi_{\uparrow}^{+}=-\Delta(x)\Psi_{\uparrow}^{+}+g_{1D}n_{\uparrow}(x)\Psi_{\downarrow},\end{equation}
 where we have defined an order parameter $\Delta(x)=-g_{1D}\langle\Psi_{\downarrow}(x)\Psi_{\uparrow}(x)\rangle$
and densities $n_{\sigma}(x)=\langle\Psi_{\sigma}^{+}(x)\Psi_{\sigma}(x)\rangle$.
The above decoupling thus yields,

\begin{eqnarray}
i\hbar\frac{\partial\Psi_{\uparrow}}{\partial t} & = & \left[{\cal H}_{\uparrow}^{s}-\mu_{\uparrow}\right]\Psi_{\uparrow}-\Delta(x)\Psi_{\downarrow}^{+},\\
i\hbar\frac{\partial\Psi_{\downarrow}}{\partial t} & = & \left[{\cal H}_{\downarrow}^{s}-\mu_{\downarrow}\right]\Psi_{\downarrow}+\Delta(x)\Psi_{\uparrow}^{+},\end{eqnarray}
 where ${\cal H}_{\sigma}^{s}=-\hbar^{2}\nabla^{2}/\left(2m\right)+g_{1D}n_{\bar{\sigma}}\left(x\right)$.
We solve the equation of motion by inserting the standard Bogoliubov
transformation: \begin{eqnarray}
\Psi_{\uparrow} & = & \mathop{\textstyle \sum}\limits _{\eta}[u_{\eta\uparrow}\left(x\right)c_{\eta\uparrow}e^{-iE_{\eta\uparrow}t/\hbar}+v_{\eta\downarrow}^{*}\left(x\right)c_{\eta\downarrow}^{+}e^{iE_{\eta\downarrow}t/\hbar}],\nonumber \\
\Psi_{\downarrow}^{+} & = & \mathop{\textstyle \sum}\limits _{\eta}[u_{\eta\downarrow}^{*}\left(x\right)c_{\eta\downarrow}^{+}e^{iE_{\eta\downarrow}t/\hbar}-v_{\eta\uparrow}\left(x\right)c_{\eta\uparrow}e^{-iE_{\eta\uparrow}t/\hbar}].\label{Bt}\end{eqnarray}
 This gives rise to the well-known BdG equations for the Bogoliubov
quasiparticle \citep{bdg},

\begin{equation}
\left[\begin{array}{cc}
{\cal H}_{\sigma}^{s}-\mu_{\sigma} & -\Delta(x)\\
-\Delta^{*}(x) & -{\cal H}_{\bar{\sigma}}^{s}+\mu_{\bar{\sigma}}\end{array}\right]\left[\begin{array}{c}
u_{\eta\sigma}\\
v_{\eta\sigma}\end{array}\right]=E_{\eta\sigma}\left[\begin{array}{c}
u_{\eta\sigma}\\
v_{\eta\sigma}\end{array}\right],\,\end{equation}
 where the wave functions $u_{\eta\sigma}\left(x\right)$ and $v_{\eta\sigma}\left(x\right)$
are normalized by \begin{equation}
\int dx\left[\left|u_{\eta\sigma}\left(x\right)\right|^{2}+\left|v_{\eta\sigma}\left(x\right)\right|^{2}\right]=1,\end{equation}
 and $E_{\eta\sigma}$ is the corresponding excitation energy.

We note that the unequal chemical potentials of spin states in the
BdG equations break the particle-hole symmetry. This leads to different
quasiparticle wave functions for the two components. However, one
may easily identify a one to one correspondence between the solution
for the spin up and spin down energy levels, \textit{i.e.}, \begin{equation}
E_{\eta\sigma}\leftrightarrow-E_{\eta\bar{\sigma}},\end{equation}
 and \begin{equation}
\left[\begin{array}{c}
u_{\eta\sigma}\left(x\right)\\
v_{\eta\sigma}\left(x\right)\end{array}\right]\leftrightarrow\left[\begin{array}{c}
-v_{\eta\bar{\sigma}}^{*}\left(x\right)\\
+u_{\eta\bar{\sigma}}^{*}\left(x\right)\end{array}\right].\end{equation}
 Because of this symmetry of the BdG equations, therefore, we may
consider the spin up part only. Letting $u_{\eta}\left(x\right)=u_{\eta\uparrow}\left(x\right)$
and $v_{\eta}\left(r\right)=v_{\eta\uparrow}\left(x\right)$, we then
remove the spin index in the equations,

\begin{equation}
\left[\begin{array}{cc}
{\cal H}_{\uparrow}^{s}-\mu_{\uparrow} & -\Delta(x)\\
-\Delta^{*}(x) & -{\cal H}_{\downarrow}^{s}+\mu_{\downarrow}\end{array}\right]\left[\begin{array}{c}
u_{\eta}\left(x\right)\\
v_{\eta}\left(x\right)\end{array}\right]=E_{\eta}\left[\begin{array}{c}
u_{\eta}\left(x\right)\\
v_{\eta}\left(x\right)\end{array}\right],\label{BdG}\end{equation}
 The order parameter $\Delta(x)$ and the linear number densities
$n_{\sigma}\left(x\right)$ should be determined self-consistently,
according to their definitions, respectively, \begin{eqnarray}
n_{\uparrow}\left(x\right) & = & \sum_{\eta}u_{\eta}^{*}(x)u_{\eta}(x)f(E_{\eta}),\label{bdg-nup}\\
n_{\downarrow}\left(x\right) & = & \sum_{\eta}v_{\eta}^{*}(x)v_{\eta}(x)f(-E_{\eta}),\label{bdg-ndw}\\
\Delta\left(x\right) & = & -g_{1D}\sum_{\eta}u_{\eta}(x)v_{\eta}^{*}(x)f(E_{\eta}).\label{bdg-gap}\end{eqnarray}
 where the summation runs over all the energy levels, including these
with negative energies $E_{\eta}<0$.

We note also that the single-plane-wave approximation described in
the last section can be recovered by replacing the level index {}``$\eta$''
with a wave vector $k$, and approximating, \begin{eqnarray}
u_{\eta}(x) & = & \bar{u}_{k}\exp\left[+i\left(\frac{q}{2}+k\right)x\right],\\
v_{\eta}(x) & = & \bar{v}_{k}\exp\left[-i\left(\frac{q}{2}-k\right)x\right],\\
E_{\eta} & = & \bar{E}_{k},\end{eqnarray}
 so that the order parameter reduces to \begin{equation}
\Delta(x)=-g_{1D}\mathop{\textstyle \sum}_{k}\bar{u}_{k}\bar{v}_{k}f(\tilde{E}_{k})\exp[iqx]=\Delta\exp[iqx],\end{equation}
 and the BdG equations become, \begin{equation}
\left[\begin{array}{cc}
\epsilon_{q/2+k}-\tilde{\mu}_{\uparrow} & -\Delta\\
-\Delta & -\epsilon_{q/2-k}+\tilde{\mu}_{\downarrow}\end{array}\right]\left[\begin{array}{c}
\bar{u}_{k}\\
\bar{v}_{k}\end{array}\right]=\bar{E}_{k}\left[\begin{array}{c}
\bar{u}_{k}\\
\bar{v}_{k}\end{array}\right],\end{equation}
 where as before, we have used the notations $\tilde{\mu}_{\uparrow}=\mu_{\uparrow}-g_{1D}n_{\downarrow}$
and $\tilde{\mu}_{\downarrow}=\mu_{\downarrow}-g_{1D}n_{\uparrow}$.
Apparently, there are two branch solutions for the quasiparticle energy
$E_{k}^{+}=(\epsilon_{q/2+k}-\epsilon_{q/2-k})/2-\delta\tilde{\mu}+E_{k}$
and $E_{k}^{-}=(\epsilon_{q/2+k}-\epsilon_{q/2-k})/2-\delta\tilde{\mu}-E_{k}$,
with the corresponding quasiparticle wave functions, \begin{equation}
\left(\begin{array}{c}
\bar{u}_{k}\\
\bar{v}_{k}\end{array}\right)_{\bar{E}_{k}=E_{k}^{+}}=\left(\begin{array}{c}
u_{k}\\
v_{k}\end{array}\right)=\Phi_{k}^{+},\end{equation}
 and \begin{equation}
\left(\begin{array}{c}
\bar{u}_{k}\\
\bar{v}_{k}\end{array}\right)_{\bar{E}_{k}=E_{k}^{-}}=\left(\begin{array}{c}
-v_{k}^{*}\\
u_{k}^{*}\end{array}\right)=\Phi_{k}^{-},\end{equation}
 respectively, exactly the same as in Eqs. (\ref{spw-energies}) and
(\ref{spw-wvfs}). Accordingly, the linear densities take the form,
\begin{eqnarray}
n_{\uparrow}\left(x\right) & = & \sum_{k}u_{k}^{2}f\left(E_{k}^{+}\right)+\sum_{k}v_{k}^{2}f\left(E_{k}^{-}\right),\\
n_{\downarrow}\left(x\right) & = & \sum_{k}v_{k}^{2}f\left(-E_{k}^{+}\right)+\sum_{k}u_{k}^{2}f\left(-E_{k}^{-}\right),\end{eqnarray}
 which turn out to be position independent due to the plane-wave form
of the wave functions.

\subsection{Hybrid BdG strategy}

We apply the above BdG formalism to a uniform Fermi gas with finite
atoms. To this end, we consider a gas of $N$ fermions in a box of
length $L$ using periodic boundary conditions, \textit{i.e.}, the
underlying wavefunction $\varphi\left(x\right)$ satisfies $\varphi\left(x=+L/2\right)=\varphi\left(x=-L/2\right)$.
The small boundary effect due to the finite size of $L$ could be
weakened or removed by enlarging the value of $L$.

In any practical calculation, because of the computational limitations,
the summation over the quasiparticle energy levels in Eqs. (\ref{bdg-nup}),
(\ref{bdg-ndw}) and (\ref{bdg-gap}) must be truncated. We therefore
following the idea of Reidl \textit{et al.} \citep{reidl} develop
a hybrid approach with the introduction of a high-energy cut-off $E_{c}$,
below which we solve the discrete BdG equations. Above the cut-off,
we use a semiclassical plane-wave approximation for the wavefunctions,
which should work well for sufficiently high-lying states.

The first step toward solving the discrete BdG equations is to assume
a real order parameter $\Delta(x)$ and then expand the quasiparticle
wavefunctions $u\left(x\right)$ and $v\left(x\right)$ using a complete
basis of single particle wavefunctions in the box $\varphi_{n}(x)$
with energy levels $\epsilon_{n}$ ($n=0,1,2,...$), \textit{i.e.},
\begin{eqnarray}
u\left(x\right) & = & \sum_{n}A_{n}\varphi_{n}(x),\\
v\left(x\right) & = & \sum_{n}B_{n}\varphi_{n}(x).\end{eqnarray}
 For the case of periodic boundary condition, we take \begin{equation}
\varphi_{n}(x)=\left\{ \begin{array}{c}
\sqrt{2/L}\cos\left[n\pi x/L\right],\text{ if }n\text{ is even;}\\
\sqrt{2/L}\sin\left[\left(n+1\right)\pi x/L\right],\text{ if }n\text{ is odd;}\end{array}\right.,\end{equation}
 and \begin{equation}
\epsilon_{n}=\left\{ \begin{array}{c}
\hbar^{2}\pi^{2}n^{2}/\left(2mL^{2}\right),\text{ if }n\text{ is even;}\\
\hbar^{2}\pi^{2}\left(n+1\right)^{2}/\left(2mL^{2}\right),\text{ if }n\text{ is odd;}\end{array}\right..\end{equation}
 The solution of the BdG equations then becomes a matrix diagonalization
problem, \begin{eqnarray}
\left[\begin{array}{cc}
{\cal H}_{nn^{\prime}}^{0\uparrow}+{\cal M}_{nn^{\prime}}^{\uparrow} & -\Delta_{nn^{\prime}}\\
-\Delta_{nn^{\prime}} & -{\cal H}_{nn^{\prime}}^{0\downarrow}-{\cal M}_{nn^{\prime}}^{\downarrow}\end{array}\right]\left[\begin{array}{c}
A_{n^{\prime}}\\
B_{n^{\prime}}\end{array}\right] & = & E\left[\begin{array}{c}
A_{n}\\
B_{n}\end{array}\right],\nonumber \\
 &  & \,\label{bdg-matrix}\end{eqnarray}
 where the matrix elements, \begin{eqnarray}
{\cal H}_{nn^{\prime}}^{0\sigma} & = & \left(\epsilon_{n}-\mu_{\sigma}\right)\delta_{nn^{\prime}},\\
{\cal M}_{nn^{\prime}}^{\sigma} & = & g_{1D}{\textstyle \int\limits _{-L/2}^{+L/2}dx\varphi_{n}(x)n_{\bar{\sigma}}\left(x\right)\varphi_{n^{\prime}}(x),\text{ }}\\
\Delta_{nn^{\prime}} & = & {\textstyle \int\limits _{-L/2}^{+L/2}dx\varphi_{n}(x)\Delta\left(x\right)\varphi_{n^{\prime}}(x).}\end{eqnarray}
 The coefficients of the eigenstate has to satisfy the condition $\sum_{n}\left(A_{n}^{2}+B_{n}^{2}\right)=1$
due to the normalization of the quasiparticle wavefunctions, \textit{i.e.},
$\int_{-L/2}^{+L/2}dx\left[u^{2}(x)+v^{2}(x)\right]=1$.

These discrete spectra (labeled by an index {}``$\eta$'') contribute
to the linear densities and the order parameter as follows, \begin{eqnarray}
n_{\uparrow d}\left(x\right) & = & \sum_{\left|E_{\eta}\right|<E_{c}}u_{\eta}^{*}(x)u_{\eta}(x)f(E_{\eta}),\\
n_{\downarrow d}\left(x\right) & = & \sum_{\left|E_{\eta}\right|<E_{c}}v_{\eta}^{*}(x)v_{\eta}(x)f(-E_{\eta}),\\
\Delta_{d}\left(x\right) & = & -g_{1D}\sum_{\left|E_{\eta}\right|<E_{c}}u_{\eta}(x)v_{\eta}^{*}(x)f(E_{\eta}),\end{eqnarray}
 where the subscript {}``$d$'' refers to the discrete levels.

On the other hand, for the high-lying states we take the semiclassical
approximation \citep{reidl}, \begin{eqnarray}
u_{\eta}(x) & \rightarrow & u(k,x)\exp\left[ikx\right],\\
v_{\eta}(x) & \rightarrow & v(k,x)\exp\left[ikx\right],\\
E_{\eta} & \rightarrow & E(k),\end{eqnarray}
 where we have regarded the wavefunctions locally at position $x$
as plane waves, whose amplitudes $u(k,x)$ and $v(k,x)$ are normalized
according to $u^{2}(k,x)+v^{2}(k,x)=1$. Keeping the most important
pair correlation terms only, it is straightforward to show that at
low temperatures, \begin{eqnarray}
n_{\uparrow c}\left(x\right) & = & \mathop{\textstyle \sum}_{k}\left[\frac{1}{2}-\frac{\epsilon_{k}-\mu}{2E_{k}(x)}\right]\Theta\left[E_{k}(x)+\delta\mu-E_{c}\right],\\
n_{\downarrow c}\left(x\right) & = & \mathop{\textstyle \sum}_{k}\left[\frac{1}{2}-\frac{\epsilon_{k}-\mu}{2E_{k}(x)}\right]\Theta\left[E_{k}(x)-\delta\mu-E_{c}\right],\\
\Delta_{c}\left(x\right) & = & -g_{1D}\mathop{\textstyle \sum}_{k}\frac{\Delta\left(x\right)}{2E_{k}(x)}\Theta\left[E_{k}(x)+\delta\mu-E_{c}\right],\end{eqnarray}
 where $E_{k}(x)=\sqrt{(\epsilon_{k}-\mu)^{2}+\Delta^{2}\left(x\right)}$
and the subscript {}``$c$'' means the continuous contribution from
high-energy levels.

The discrete and continuous parts of the order parameter may be combined
together to give, \begin{equation}
\Delta\left(x\right)=-g_{1D}^{eff}\left(x\right)\sum_{\left|E_{\eta}\right|<E_{c}}u_{\eta}(x)v_{\eta}^{*}(x)f(E_{\eta}),\label{gapeff}\end{equation}
 where we have defined a position dependent effective 1D coupling
constant $g_{1D}^{eff}\left(x\right)$, which satisfies, \begin{equation}
\frac{1}{g_{1D}^{eff}\left(x\right)}=\frac{1}{g_{1D}}+g(x),\label{g1deff}\end{equation}
 where \begin{equation}
g\left(x\right)=\sum_{k}\frac{1}{2E_{k}(x)}\Theta\left[E_{k}(x)+\delta\mu-E_{c}\right].\end{equation}
 The summation over the momentum $k$ may be converted into a continuous
integral of the energy. As a result, we obtain, \begin{eqnarray}
n_{\uparrow c}\left(x\right) & = & \frac{\left(2m\right)^{1/2}}{4\pi\hbar}\int\limits _{E_{c}}^{\infty}d\epsilon\left[\frac{\epsilon-\delta\mu}{\sqrt{\left(\epsilon-\delta\mu\right)^{2}-\Delta^{2}\left(x\right)}}-1\right]\nonumber \\
 &  & \times\frac{1}{\left[\mu+\sqrt{\left(\epsilon-\delta\mu\right)^{2}-\Delta^{2}\left(x\right)}\right]^{1/2}},\\
n_{\downarrow c}\left(x\right) & = & \frac{\left(2m\right)^{1/2}}{4\pi\hbar}\int\limits _{E_{c}}^{\infty}d\epsilon\left[\frac{\epsilon+\delta\mu}{\sqrt{\left(\epsilon+\delta\mu\right)^{2}-\Delta^{2}\left(x\right)}}-1\right]\nonumber \\
 &  & \times\frac{1}{\left[\mu+\sqrt{\left(\epsilon+\delta\mu\right)^{2}-\Delta^{2}\left(x\right)}\right]^{1/2}},\end{eqnarray}
 and \begin{eqnarray}
g(x) & = & \frac{\left(2m\right)^{1/2}}{4\pi\hbar}\int\limits _{E_{c}}^{\infty}d\epsilon\frac{1}{\sqrt{\left(\epsilon-\delta\mu\right)^{2}-\Delta^{2}\left(x\right)}}\nonumber \\
 &  & \times\frac{1}{\left[\mu+\sqrt{\left(\epsilon-\delta\mu\right)^{2}-\Delta^{2}\left(x\right)}\right]^{1/2}}.\label{gx}\end{eqnarray}

We can now summarize the entire procedure used to obtain the BdG solutions.
The key step is to solve the eigenvalue problem (\ref{bdg-matrix}).
As the calculation of matrix elements involves the order parameter
and linear densities that are yet to be determined, a self-consistent
iterative procedure is required. For a given number of atoms ($N=N_{\uparrow}+N_{\downarrow}$
and $\delta N=N_{\uparrow}-N_{\downarrow}$), temperature and interaction
coupling $g_{1D}$, we:

\begin{description}
\item [{{(a)}}] start with an initial guess or a previously determined
better estimate for $\Delta\left(x\right)$,
\item [{{(b)}}] solve Eqs. (\ref{g1deff}) and (\ref{gx}) for the effective
coupling constant,
\item [{{(c)}}] then solve Eq. (\ref{bdg-matrix}) for all the quasiparticle
wavefunctions up to the chosen energy cut-off to find $u_{\eta}\left(x\right)$
and $v_{\eta}\left(x\right)$, and finally determine an improved value
for the order parameter from Eq. (\ref{gapeff}).
\end{description}
During the iteration, the density profiles $n_{\uparrow}(x)=n_{\uparrow d}(x)+n_{\uparrow c}(x)$
and $n_{\downarrow}(x)=n_{\downarrow d}(x)+n_{\downarrow c}(x)$ are
updated. The chemical potentials $\mu$ and $\delta\mu$ are also
adjusted slightly in each iterative step to enforce the number-conservation
condition that $\int_{-L/2}^{+L/2}dx[n_{\uparrow}(x)+n_{\downarrow}(x)]{\bf =}N$
and $\int_{-L/2}^{+L/2}dx[n_{\uparrow\text{ }}(x)-n_{\downarrow\text{ }}(x)]{\bf =}\delta N$,
until final convergence is reached.

\subsection{The structure of FFLO states}

Using the self-consistent BdG formalism we can work out the detailed
structure of mean-field or FFLO states. To make the equations dimensionless,
as before we take the Fermi wave vector $k_{F}=\pi n/2=\pi N/(2L)$
and the Fermi energy $\epsilon_{F}=\hbar^{2}k_{F}^{2}/(2m)$ as the
units of the momentum and energy, respectively, \textit{i.e.}, by
setting $\hbar=1$ and $2m=1$, and $k_{F}=1$. Therefore, the size
of the box $L=\pi N/2$ can be enlarged by increasing the number of
total atoms $N$. In the following calculations, we use $N=200$,
which in most cases we find is large enough to effectively minimize
the boundary effects. Further, we take a cut-off energy $E_{c}=16\epsilon_{F}$.
This cut-off energy is already sufficient large because of the high
efficiency of our hybrid strategy. Accordingly, we setup a set of
single-particle-state basis $\varphi_{n}(x)$, with the highest energy
level larger than the cut-off energy.

The initial guess for the order parameter $\Delta(x)$ could be arbitrary.
However, we find that in general there are many locally metastable
solutions after the iteration, which can be classified uniquely by
their periodicity. This is due to the existence of the periodic boundary
condition that requires that the order parameter should be a periodic
function of length $L/n$, where $n$ is an integer. We therefore
compare the energy (or free energy) of the solutions with different
periodicity, and select the one with the lowest energy as the ground
state.

%
\begin{figure}
\begin{centering}
\includegraphics[clip,width=0.45\textwidth]{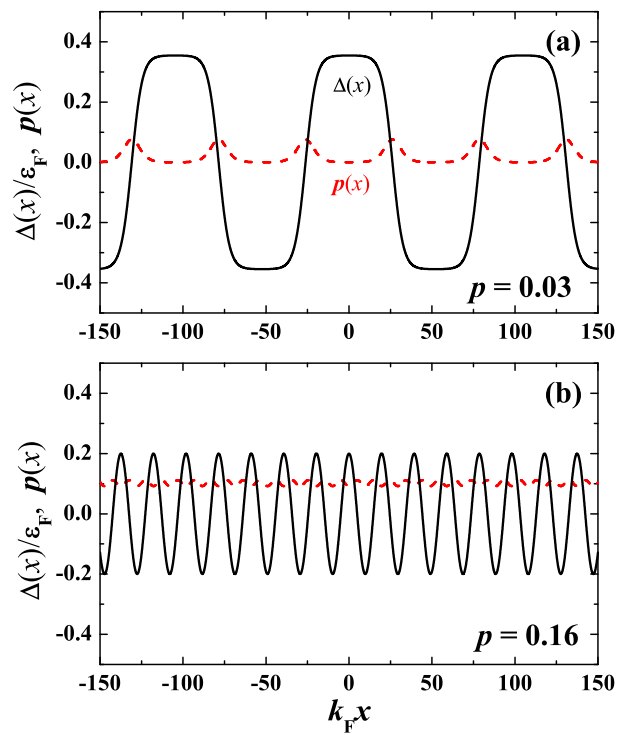}
\par\end{centering}

\caption{(Color online) Spatial structures of the inhomogeneous FFLO states
at an interaction strength $\gamma=1.6$ and at two spin polarizations
as indicated. The calculations have been done for a uniform gas confined
in a box, using the self-consistent BdG equations. The solid line
and the dashed line refer to the order parameter and the local spin
polarization, respectively.}

\label{fig7}
\end{figure}


We present in Fig. 7 the spatial distribution of the order parameter
$\Delta\left(x\right)$ and the local spin polarization \begin{equation}
p\left(x\right)=\frac{n_{\uparrow}\left(x\right)-n_{\downarrow}\left(x\right)}{n_{\uparrow}\left(x\right)+n_{\downarrow}\left(x\right)}\end{equation}
 for a uniform Fermi gas with total polarization $p=0.03$ (a) and
$p=0.16$ (b) at a typical coupling constant $\gamma=1.6$. The most
notable feature of the figure is that at a small total polarization
(Fig. 7a), the order parameter switches between two values: $+\Delta_{0}$
and $-\Delta_{0}$, where $\Delta_{0}$ is the full gap of an unpolarized
gas at the same coupling. Many instantons and anti-instantons (or
kinks and anti-kinks) then appear and carry the excess spin up (majority)
atoms since the local polarization $p(x)$ shows pronounced peaks
right at the position where the order parameter vanishes. These features
are not unlike a phase separation, except that a regular, periodic
domain structure is obtained. Thus, in the limit of small polarization,
the order parameter may be viewed as an instanton gas, with the number
of instantons roughly proportional to the spin polarization. Within
this picture, we anticipate that an FFLO state emerges as soon as
the polarization becomes nonzero. In contrast, for a large total polarization
(Fig. 7b), the order parameter is well approximated by a cosine function,
as expected earlier by Larkin and Ovchinnikov. It is a superposition
of two single-plane-waves going in opposite directions, with a much
reduced amplitude compared to $\Delta_{0}$.

We note that in the weak coupling limit, a snoidal solution of the
order parameter for the BdG equations was found analytically if one
linearizes the single particle spectrum at the Fermi surface \citep{mf1d,buzdin},
which gives qualitatively the same behavior as shown in Fig. 7.

\subsection{Phase diagram from BdG solutions}

%
\begin{figure}
\begin{centering}
\includegraphics[clip,width=0.45\textwidth]{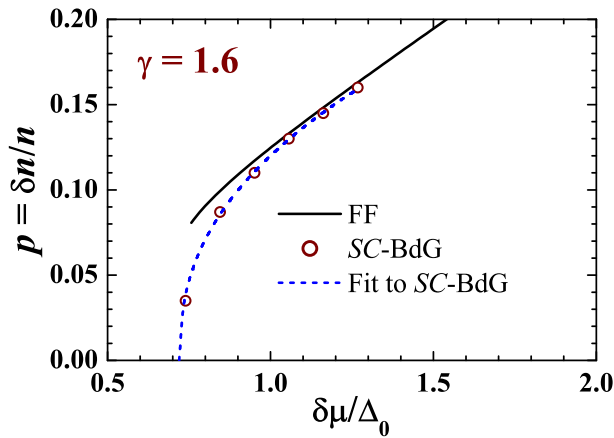}
\par\end{centering}

\caption{(Color online) Spin polarization versus the chemical potential difference
at an interaction strength $\gamma=1.6$, obtained from the single-plane-wave
approximation (solid line) and the self-consistent BdG calculations
(open circles). While the spin polarization in the FF state shows
a jump as a function of the chemical potential difference, the more
accurate self-consistent BdG prediction suggests that the spin polarization
emerges from zero continuously with increasing the chemical potential
difference. The dashed line is a power-law fit to the self-consistent
BdG results.}

\label{fig8}
\end{figure}


We examine the phase diagram obtained by the single-plane-wave approximation
(Fig. 5). For this purpose, we compare the results of the spin polarization
versus the chemical potential difference, as predicted respectively
by the self-consistent BdG formalism and the single-plane-wave approximation
or the FF solution. As shown in Fig. 8, the self-consistent prediction
agrees very well with that of the FF solution at a large chemical
potential difference. However, approaching to the BCS-FFLO transition
point, they differ largely. The quick fall of the spin polarization
in the self-consistent BdG indicates strongly the existence of a FFLO
state with an arbitrary small spin polarization. As the spin polarization
is a first order derivative of the energy, this is a solid evidence
for the smooth transition from the BCS state to the FFLO state. We
therefore conclude that although the single-plane-wave approximation
gives a reasonable description at the large chemical potential difference,
it does not predict the correct phase transition between BCS and FFLO
states.

We may extract the critical behavior at the transition point by numerically
analyzing the self-consistent data. Assuming a pow-law dependence
of the spin polarization on the chemical potential difference, $p\propto(\delta\mu-\delta\mu_{c})^{\alpha}$,
we find that $\alpha\simeq0.4$, in good agreement with a non-perturbative
bosonization prediction \citep{yang1d}, $\alpha=0.5$. The small
discrepancy may be caused by the use of a finite length $L$, which
becomes increasingly in-efficient due to the divergent correlation
length towards the transition point.

\subsection{Local fermionic density of states}

%
\begin{figure}
\begin{centering}
\includegraphics[clip,width=0.45\textwidth]{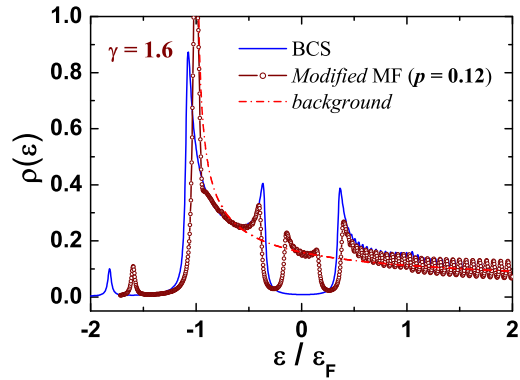}
\par\end{centering}

\caption{(Color online) Local fermionic density of states of a uniform polarized
Fermi gas at an interaction strength $\gamma=1.6$, calculated using
the self-consistent BdG equations.}

\label{fig9}
\end{figure}


We finally calculate the local density of states in the self-consistent
BdG solutions, which is given by, \begin{eqnarray}
\rho_{\uparrow}\left(x,\epsilon\right) & = & \sum_{\eta}u_{\eta}^{2}\left(x\right)\delta\left(\epsilon-E_{\eta}\right),\\
\rho_{\downarrow}\left(x,\epsilon\right) & = & \sum_{\eta}v_{\eta}^{2}\left(x\right)\delta\left(\epsilon+E_{\eta}\right).\end{eqnarray}
 In Fig. 9, we show how the local density of states at origin evolves
with increasing the spin polarization $p$ from zero to $0.12$. Here
a small spectral broadening of about $0.02\epsilon_{F}$ has been
used to regularize the delta function. We find again a nonzero density
of states at the Fermi surface for a polarized Fermi gas, contributed
by the mid-gap states. As a result, the original BCS gap of a width
$2\Delta_{0}$ is split into two sub-gaps with a much smaller width.

\section{Exact Bethe ansatz solution in a homogeneous gas}

The validity of mean-field results in 1D is not immediately clear,
as pair fluctuations become increasingly important in lower dimensions.
Fortunately, without the trap the Hamiltonian (\ref{model}) of a
free polarized Fermi gas is exactly soluble, using the Bethe ansatz
technique \citep{gaudin,takahashi}. We therefore can use the exact
solution as a benchmark to test the validity of various mean-field
approaches.

In the thermodynamic limit, the ground state of a homogeneous gas
with fixed linear densities $n_{\uparrow}$ and $n_{\downarrow}$
may be obtained from a set of Gaudin integral equations \citep{takahashi},
\begin{eqnarray}
\pi\rho\left(k\right) & = & \frac{1}{2}-{\textstyle \int\limits _{-B}^{B}\frac{c^{\prime}\sigma\left(\Lambda\right)d\Lambda}{c^{\prime2}+\left(k-\Lambda\right)^{2}},}\\
\pi\sigma\left(\Lambda\right) & = & 1-{\textstyle \int\limits _{-Q}^{Q}\frac{c^{\prime}\rho\left(k\right)dk}{c^{\prime2}+\left(\Lambda-k\right)^{2}}-{\textstyle \int\limits _{-B}^{B}\frac{c\sigma\left(\Lambda^{\prime}\right)d\Lambda^{\prime}}{c^{2}+\left(\Lambda-\Lambda^{\prime}\right)^{2}},\,}}\end{eqnarray}
 and \begin{eqnarray}
\epsilon_{gs} & = & \frac{\hbar^{2}}{2m}\left[{\textstyle \int\limits _{-Q}^{Q}k^{2}\rho\left(k\right)+{\textstyle \int\limits _{-B}^{B}2\Lambda^{2}\sigma\left(\Lambda\right)-\frac{n_{\downarrow}c^{2}}{2}}}\right],\nonumber \\
n_{\uparrow}-n_{\downarrow} & = & {\textstyle \int_{-Q}^{Q}\rho(k)dk,}\\
n_{\downarrow} & = & {\textstyle \int_{-B}^{B}\sigma(\Lambda)d\Lambda,}\end{eqnarray}
 where $\epsilon_{gs}$ is the ground state energy density, the couplings
$c=n\gamma$ and $c^{\prime}=c/2$. The functions $\rho(k)$ and $\sigma(\Lambda)$
are, respectively, the quasi-momentum distributions with the cut-off
rapidities $Q$ and $B$ to be determined by the normalization condition
for $\delta n=n_{\uparrow}-n_{\downarrow}$ and $n_{\downarrow}$.
The last term in $\epsilon_{gs}$ is simply the contribution from
$n_{\downarrow}$ paired two-fermion bound states with binding energy
\begin{equation}
\epsilon_{b}=\frac{\hbar^{2}c^{2}}{4m}=\frac{\hbar^{2}}{ma_{1D}^{2}}.\end{equation}
 The chemical potential and the chemical potential difference can
be obtained by $\mu=\partial\epsilon_{gs}/\partial n$ and $\delta\mu=\partial\epsilon_{gs}/\partial\delta n,$
respectively.

\subsection{Gaudin solutions}

The Gaudin integral equations have to be solved numerically for a
general spin polarization $p=\delta n/n$. To do so, we introduce
two new variables $x=k/Q$ and $y=\Lambda/B$, and rewrite the quasi-momentum
distribution functions, \begin{eqnarray}
g_{c}\left(x\right) & = & \rho\left(Qx\right)=\rho\left(k\right),\\
g_{s}\left(y\right) & = & \sigma\left(By\right)=\sigma\left(\Lambda\right).\end{eqnarray}
 Further, the two cut-off rapidities may be represented by, respectively,
$Q=n\gamma/\lambda_{c}$ and $B=n\gamma/\lambda_{s}$. In such a way,
the Gaudin integral equations can be rewritten in a dimensionless
form, \begin{eqnarray}
g_{c}\left(x\right) & = & \frac{1}{2\pi}-{\textstyle \int\limits _{-1}^{+1}\frac{g_{s}\left(y\right)/\left[2\pi\lambda_{s}\right]}{\frac{1}{4}+\left(\frac{x}{\lambda_{c}}-\frac{y}{\lambda_{s}}\right)^{2}}dy,}\\
g_{s}\left(x\right) & = & \frac{1}{\pi}-{\textstyle \int\limits _{-1}^{+1}\frac{g_{c}\left(y\right)/\left[2\pi\lambda_{c}\right]}{\frac{1}{4}+\left(\frac{x}{\lambda_{s}}-\frac{y}{\lambda_{c}}\right)^{2}}dy-{\textstyle \int\limits _{-1}^{+1}\frac{g_{s}\left(y\right)/\left[\pi\lambda_{s}\right]}{1+\left(\frac{x-y}{\lambda_{s}}\right)^{2}}dy,}}\end{eqnarray}
 together with the normalization conditions, \begin{eqnarray}
\lambda_{c} & = & \frac{\gamma}{p}{\textstyle \int\limits _{-1}^{+1}g_{c}\left(x\right)dx,}\\
\lambda_{s} & = & \frac{2\gamma}{1-p}{\textstyle \int\limits _{-1}^{+1}g_{s}\left(x\right)dx.}\end{eqnarray}
 Numerically, the dimensionless integral equations have been solved
by decomposing the integrals on a grid with $N$ points $\{x_{i};x_{i}\in\left[-1,+1\right]\}$.
In detail, we start from a set of trial distributions $g_{c}^{(0)}(x_{i})$
and $g_{s}^{(0)}(x_{i})$, and the corresponding parameters of $\lambda_{c}^{(0)}$
and $\lambda_{s}^{(0)}$. Following the standard method for the integrals
\citep{lieb}, we obtain $g_{c}(x_{i})$ and $g_{s}(x_{i})$. Let
$g_{c}^{(1)}(x_{i})=\alpha g_{c}^{(0)}(x_{i})+(1-\alpha)g_{c}(x_{i})$
and $g_{s}^{(1)}(x_{i})=\alpha g_{s}^{(0)}(x_{i})+(1-\alpha)g_{s}(x_{i})$
(where $\alpha$ is a positive real number between $0$ and $1$,
depending the value of the spin polarization) be the new trial distributions,
and update $\lambda_{c}^{(1)}$ and $\lambda_{s}^{(1)}$ accordingly.
Repeat the above procedure until $g_{c}(x_{i})$ and $g_{s}(x_{i})$
agree with their trial distributions within a certain range. Then,
the energy density \begin{equation}
\epsilon_{gs}=\frac{\hbar^{2}n^{3}}{2m}e\left(\gamma,p\right)-n_{\downarrow}\epsilon_{b}\end{equation}
 is calculated by: \begin{equation}
e\left(\gamma,p\right)=\frac{\gamma^{3}}{\lambda_{c}^{3}}{\textstyle \int\limits _{-1}^{+1}x^{2}g_{c}\left(x\right)dx+\frac{\gamma^{3}}{\lambda_{s}^{3}}{\textstyle \int\limits _{-1}^{+1}2x^{2}g_{s}\left(x\right)dx.}}\end{equation}
 We find that this iterative method for solving the Gaudin integral
equations is very stable. The chemical potential and chemical potential
difference can also be calculated accurately by a numerical derivative.

%
\begin{figure}
\begin{centering}
\includegraphics[clip,width=0.45\textwidth]{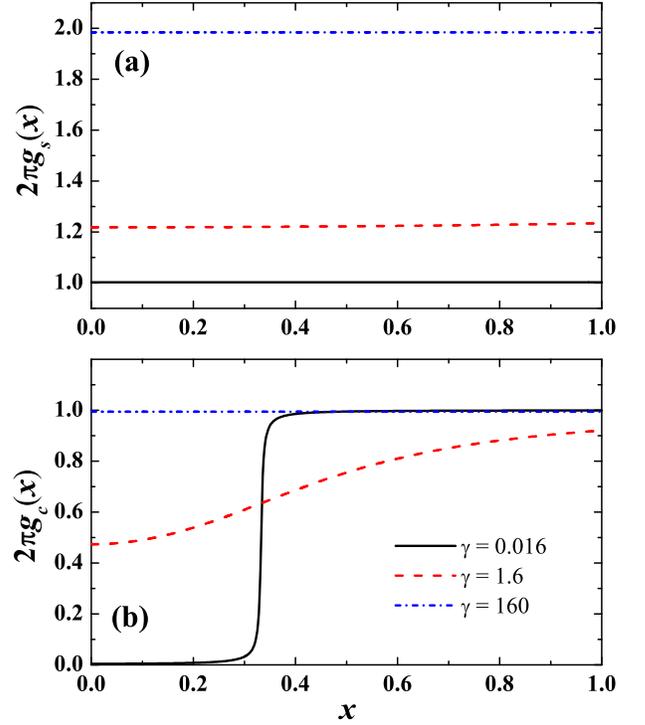}
\par\end{centering}

\caption{(Color online) Gaudin solution for the dimensionless quasi-momentum
distributions at a spin polarization $p=0.5$ and at several interaction
couplings as indicated.}

\label{fig10}
\end{figure}


For an illustrative purpose, we plot in Fig. 10 the quasi-momentum
distribution functions $g_{s}(x)$ (Fig. 10a) and $g_{c}(x)$ (Fig.
10b) at a spin polarization $p=0.5$ for three interaction strengths
as indicated. As $g_{s}(x)$ and $g_{c}(x)$ are both even functions,
we show only the part with a positive $x$. For a large interaction
strength, they approach $1/\pi$ and $1/(2\pi)$ respectively. On
the other hand, for a weak interaction, $g_{s}(x)$ reduces to $1/(2\pi)$
and $g_{c}(x)$ jumps from zero to $1/(2\pi)$ at a certain value
of $x$.

%
\begin{figure}
\begin{centering}
\includegraphics[clip,width=0.45\textwidth]{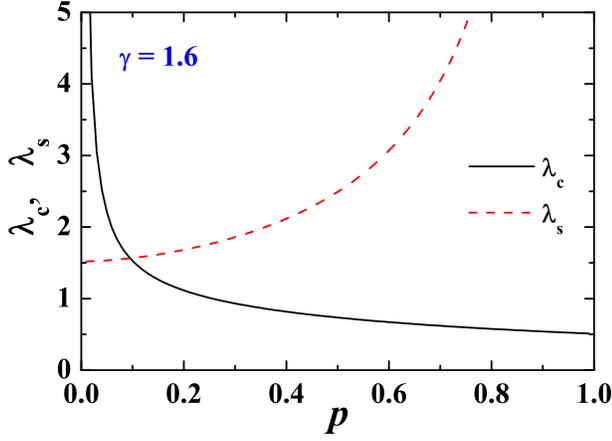}
\par\end{centering}

\caption{(Color online) Gaudin solution for the dimensionless parameters $\lambda_{c}$
and $\lambda_{s}$, as a function of the spin polarization at an interaction
strength $\gamma=1.6$.}

\label{fig11}
\end{figure}


At $\gamma=1.6$ the dimensionless parameters $\lambda_{c}$ and $\lambda_{s}$
as a function of the spin polarization are shown in Fig. 11. They
diverge respectively as $1/p$ and $1/(1-p)$ when the spin polarization
goes to $0$ or $1$.

\subsection{Analytic results in limiting cases}

The asymptotic behavior of the Gaudin solution may be obtained in
the strongly and weakly interacting limits. For a strongly interacting
gas, for which the dimensionless coupling constant $\gamma\gg1$,
the parameters $\lambda_{c}$ and $\lambda_{s}$ are sufficient large.
Therefore, the integrals in the Gaudin equations becomes extremely
small. Hence, the quasi-momentum distributions $g_{c}(x)$ and $g_{s}(x)$
are essentially constant. Expanding to the order $1/\gamma^{3}$,
we find that, \begin{eqnarray}
g_{c}(x) & = & \frac{1}{2\pi}-\frac{1-p}{\pi\gamma}+o\left(\frac{1}{\gamma^{3}}\right),\\
g_{s}(x) & = & \frac{1}{\pi}-\frac{1+3p}{2\pi\gamma}+o\left(\frac{1}{\gamma^{3}}\right).\end{eqnarray}
 It is then straightforward to show that to leading order in $1/\gamma$,
\begin{eqnarray}
e\left(\gamma,p\right) & \simeq & \frac{\pi^{2}\left(1-p\right)^{3}}{48}+\frac{\pi^{2}p^{3}}{3},\\
\mu & \simeq & -\frac{\epsilon_{b}}{2}+\frac{\hbar^{2}n^{2}}{2m}\frac{\pi^{2}\left(1-p\right)^{2}}{16},\\
\delta\mu & \simeq & \frac{\epsilon_{b}}{2}-\frac{\hbar^{2}n^{2}}{2m}\frac{\pi^{2}\left(1-p\right)^{2}}{16}+\frac{\hbar^{2}n^{2}}{2m}\pi^{2}p^{2}.\label{dmu-strong}\end{eqnarray}
 Recalling that $n_{\downarrow}=n(1-p)/2$, the chemical potential,
as well as the first two terms on the right-hand side of the chemical
potential difference, coincide in magnitude with the chemical potential
of a Tonks-Girardeau bosonic gas of paired $n_{\downarrow}$ dimers
\citep{lieb}, which is fermionized due to strong attractions. The
third term in the chemical potential difference, on the other hand,
is equal to the chemical potential of residual unpaired $n_{\uparrow}-n_{\downarrow}$
fermions. Therefore, in the strong coupling regime the polarized gas
behaves like an incoherent mixture of a molecular Bose gas and a fully
polarized single-species Fermi gas.

The analytic derivation in the weak coupling limit $\gamma\ll1$ is
much more subtle since the quasi-momentum distribution $g_{c}(x)$
contains a sharp jump whose width ($\sim\gamma$) is extremely small,
as shown in Fig. 10b for $\gamma=0.016$. However, as a leading approximation,
we may take $g_{c}(x)$ as a step function. It is then easy to show
that ($\gamma\ll\max\{p,1-p\}$), \begin{eqnarray}
g_{c}(x) & = & \left\{ \begin{array}{c}
0,\quad\left|x\right|<\left(1-p\right)/\left(1+p\right)\\
1/\left(2\pi\right),\quad\left|x\right|>\left(1-p\right)/\left(1+p\right)\end{array}\right.,\\
g_{s}(x) & = & 1/\left(2\pi\right).\end{eqnarray}
 As a result, the ground state energy density and the chemical potentials
are given by \begin{eqnarray}
e\left(\gamma,p\right) & \simeq & \frac{\pi^{2}}{12}\left(1+3p^{2}\right)-\frac{\gamma}{2}\left(1-p^{2}\right),\\
\mu & \simeq & \frac{\hbar^{2}n^{2}}{2m}\frac{\pi^{2}}{4}\left(1+p^{2}\right)+\frac{\hbar^{2}n^{2}}{2m}\gamma,\\
\delta\mu & \simeq & \frac{\hbar^{2}n^{2}}{2m}\frac{\pi^{2}}{2}p+\frac{\hbar^{2}n^{2}}{2m}\gamma p,\label{dmu-weak}\end{eqnarray}
 where the first term on the right-hand side corresponds to an ideal
polarized gas, while the second term arises from the mean-field Hartree-Fock
interactions. We note that a non-perturbative term of order $\gamma^{2}\ln\gamma$
will occur if one improves the quasi-momentum distribution functions
by explicitly taking into account the width of the jump in $g_{c}(x)$.

\subsection{Mean-field approaches versus exact solutions}

%
\begin{figure}
\begin{centering}
\includegraphics[clip,width=0.45\textwidth]{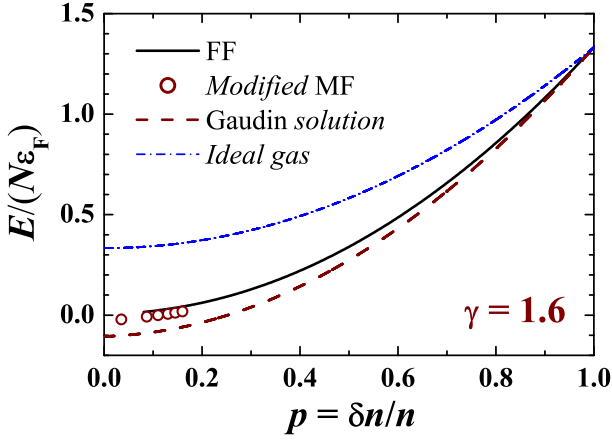}
\par\end{centering}

\caption{(Color online) Comparison of the mean-field energy to the exact results
obtained from the Bethe ansatz solution at an interaction strength
$\gamma=1.6$. For a reference, we plot also the energy of an ideal
polarized gas. Presumably, the small discrepancy between the mean-field
and exact results is due to the pair fluctuation effects.}

\label{fig12}
\end{figure}


%
\begin{figure}
\begin{centering}
\includegraphics[clip,width=0.45\textwidth]{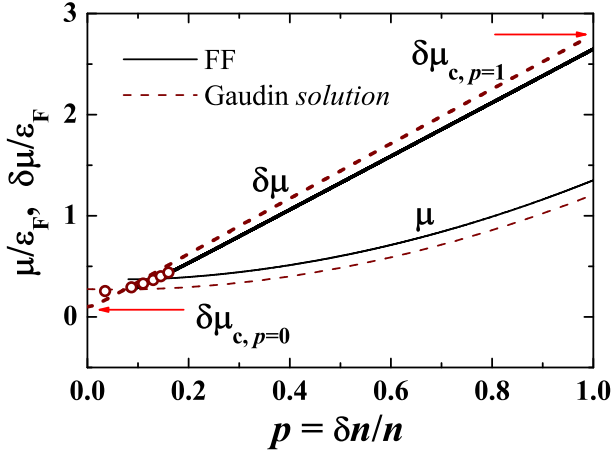}
\par\end{centering}

\caption{(Color online) Comparison of the mean-field chemical potentials to
the exact results obtained from the Bethe ansatz solution at an interaction
strength $\gamma=1.6$. The arrows point to two critical chemical
potential differences, between which a polarized superfluid exists.}

\label{fig13}
\end{figure}


We are now ready to verify the accuracy of the mean-field approaches.
In Figs. 12 and 13, we compare the energy and chemical potentials
of the exact Gaudin solutions with that from mean-field calculations,
with either a single-plane-wave like (labeled as {}``FF'') or a
self-consistently determined (denoted by {}``\textit{SC}-BdG'')
order parameter. For comparison, the energy of an ideal polarization
gas is also shown. For a moderate interaction coupling $\gamma=1.6$,
we find a reasonable agreement. The residual discrepancy could be
ascribed to pair fluctuations, which are small but not negligible.
We have also checked that the agreement becomes increasingly better
(as expected), with decreasing interaction strength. With these observations,
we therefore confirm the validity of the mean-field theories for the
weakly and moderately interacting regimes.

On the other hand, the good agreement between the Gaudin solutions
and the mean-field results suggests strongly that the partially polarized
solution found in the exact Bethe ansatz method is of FFLO character.
We note that a calculation of the nonlocal pair correlation functions
in the exact solution would be very useful to unambiguously determine
its structure. However, this is extremely difficult due to the complicated
ground state wavefunctions from the Bethe ansatz.

\subsection{Quantitative phase diagram of a homogeneous polarized Fermi gas}

%
\begin{figure}
\begin{centering}
\includegraphics[clip,width=0.45\textwidth]{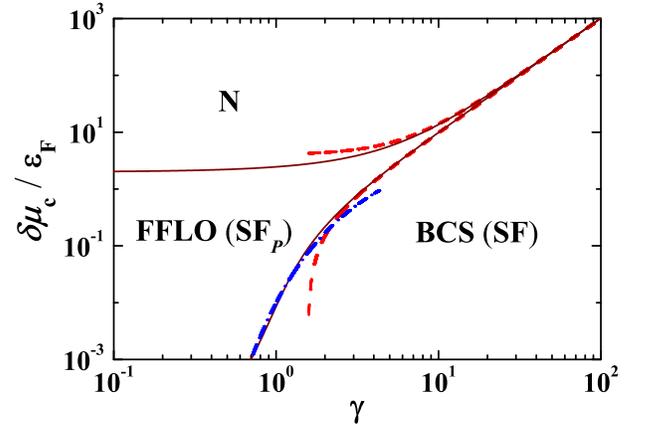}
\par\end{centering}

\caption{(Color online) Phase diagram of a one-dimensional homogeneous spin-polarized
Fermi gas. The dot-dashed line refers to the asymptotic expression
of the critical chemical potential difference in the weak coupling
limit, i.e., Eq. (\ref{critp0_weak}), while the two dashed lines
are respectively, the strong-coupling expansion of the critical chemical
potential difference, as described in Eqs. (\ref{critp0_strong})
and (\ref{critp1_strong}).}

\label{fig14}
\end{figure}


Gathering all the information from the Gaudin integral solutions and
the two mean-field results, we arrive at a quantitative phase diagram
for a homogeneous polarized Fermi gas \citep{hldprl1d,orso}. For
a given interaction strength the chemical potential difference takes
values between two thresholds, $\delta\mu_{c,p=0}$ and $\delta\mu_{c,p=1}$,
as indicated by arrows in Fig. 13 for $\gamma=1.6$. Below the first
threshold $\delta\mu_{c,p=0}$, the gas persists in the BCS-like superfluid
state with zero polarization (SF), while above the second critical
value $\delta\mu_{c,p=1}$, a fully polarized normal state appears
(N). In between, a superfluid state with finite polarization (SF$_{P}$)
is favored. As stated earlier, the SF$_{P}$ has a FFLO structure
in character. Physically $\delta\mu_{c,p=0}$ is the energy cost required
to break spin-singlet pairs in unpolarized superfluid, \textit{i.e.},
the spin gap, while $\delta\mu_{c,p=1}$ is also associated with the
pair-breaking (for the last pair), but is enhanced due to the Pauli
repulsion from existing fermions. The dependence of $\delta\mu_{c,p=0}$
and $\delta\mu_{c,p=1}$ on the parameter $\gamma$ is reported in
Fig. 14, constituting a homogeneous phase diagram.

The behavior of the critical chemical potential difference in the
weak and strong coupling limits may be worked out analytically. In
the strongly interacting regime of $\gamma\rightarrow\infty$, from
its asymptotic expression (\ref{dmu-strong}) we find that, \begin{eqnarray}
\delta\mu_{c,p=0} & \simeq & \frac{\epsilon_{b}}{2}-\frac{\hbar^{2}n^{2}}{2m}\frac{\pi^{2}}{16},\label{critp0_strong}\\
\delta\mu_{c,p=1} & \simeq & \frac{\epsilon_{b}}{2}+\frac{\hbar^{2}n^{2}}{2m}\pi^{2}.\label{critp1_strong}\end{eqnarray}
 While in the weakly interacting limit of $\gamma\rightarrow0$, only
$\delta\mu_{c,p=1}$ can be determined from the weak coupling expression
(\ref{dmu-weak}), \begin{equation}
\delta\mu_{c,p=1}\simeq\frac{\hbar^{2}n^{2}}{2m}\left(\frac{\pi^{2}}{2}+\gamma\right),\end{equation}
 as the validity of the equation is restricted to $\gamma\ll\max\{p,1-p\}$.
The determination of $\delta\mu_{c,p=0}$ as $\gamma\rightarrow0$
turns out to be very difficult. Fortunately, it has been studied by
Krivnov and Ovchinnikov \citep{krivnov}, and Fuchs, Recati, and Zwerger
\citep{zwerger} in detail. Here we only quote their result, \begin{equation}
\delta\mu_{c,p=0}\simeq\frac{\hbar^{2}n^{2}}{2m}2\sqrt{\pi\gamma}\exp\left[-\frac{\pi^{2}}{2\gamma}\right].\label{critp0_weak}\end{equation}
 This predicts the same exponent $-\pi^{2}/(2\gamma)$ as the BCS
mean-field theory. However, there is a different power-law dependence
of the prefactor on the dimensionless coupling constant. \textit{i.e.},
it has an extra $\sqrt{\gamma}$ factor. In Fig. 14, we plot these
analytic predictions using dashed and dot-dashed lines. They are in
excellent agreement with the exact numerical results in the regions
where they are valid.

%
\begin{figure}
\begin{centering}
\includegraphics[clip,width=0.45\textwidth]{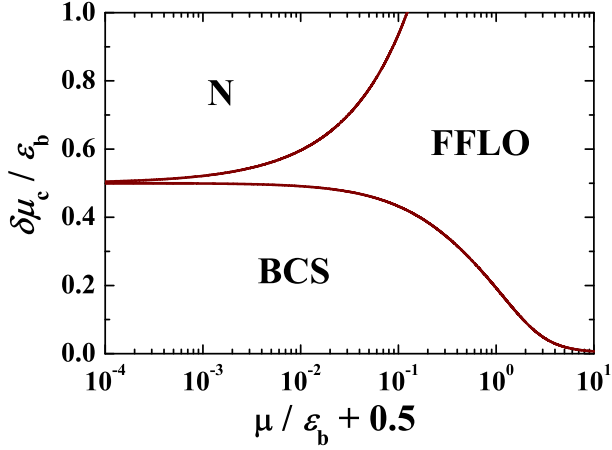}
\par\end{centering}

\caption{(Color online) Same phase diagram as in Fig. 12, but plotted here
in the plane of the chemical potential and the chemical potential
difference. Note that the chemical potential difference is in units
of the binding energy, so that the diagram is particularly useful
for the case with a fixed interaction strength, but varying densities.}

\label{fig15}
\end{figure}


For a later reference, in Fig. 15 we reconstruct the phase diagram
in the plane of the chemical potential and the chemical potential
difference. Both of them are measured in units of the binding energy.
It is clear that in the strong coupling limit, the two critical chemical
potential differences converge to the half of the binding energy,
and the phase space for the FFLO states therefore becomes much narrower.

\section{Self-consistent BdG approach in a harmonic trap}

To make a quantitative contact with the on-going experiments, it is
crucial to take into account the trapping potential that is necessary
to prevent the atoms from escaping. In this section we turn to describe
a 1D polarized gas in harmonic traps, using the mean-field BdG equations.

With the trap $V_{trap}\left(x\right)=m\omega^{2}x^{2}/2$, the BdG
formalism is essentially the same as that under a periodic boundary
condition, except a few modifications: (1) First, one has to replace
everywhere the chemical potential $\mu$ by a local potential $\mu-V_{trap}(x)$.
(2) Accordingly, to solve the BdG equation, it is convenient to use
the eigenfunctions of the harmonic trap, \begin{equation}
\varphi_{n}\left(x\right)=A_{n}H_{n}\left(\frac{x}{a_{ho}}\right)\exp\left(-\frac{x^{2}}{2a_{ho}^{2}}\right),\end{equation}
 as the set of the expanding basis. Here $H_{n}\left(x\right)$ is
the Hermite polynomial with an order $n$, $a_{ho}=\left[\hbar/\left(m\omega\right)\right]^{1/2}$
the characteristic harmonic oscillator length, and $A_{n}=\sqrt{1/(\pi^{1/2}2^{n}n!)}$
the normalization factor for single particle eigenfunctions. (3) Thirdly,
for the convenience of the numerical calculations, it is better to
take the trap units, \textit{i.e.}, $m=\hbar=\omega=1$, so that the
length and energy will be measured in units of the characteristic
harmonic oscillator length $a_{ho}$ and $\hbar\omega$, respectively.
(4) Finally, in the presence of the trap, there is no restriction
for the initial guess of the order parameter. We may then initialize
the order parameter by choosing some random values.

We have performed a calculation for a gas with $N=128$ fermions in
traps at zero temperature. The Fermi energy under the unpolarized
condition is $E_{F}=(N/2)\hbar\omega=64\hbar\omega$. We therefore
take a cut-off energy $E_{c}=6E_{F}=384\hbar\omega$ and keep up to
$6N=768$ single particle eigenfunctions. These parameters are already
very large to ensure the accuracy of the calculations. As mentioned
earlier, we use the dimensionless coupling parameter at the trap center,
$\gamma_{0}=\pi a_{ho}/(N^{1/2}a_{1D})$, to characterize the interaction.
In Fig. 16, we present the BdG results for the density profiles (solid
lines) and the order parameter (dot-dashed lines) at a moderate interaction
strength $\gamma_{0}=1.6$ for three total spin polarizations as indicated.

%
\begin{figure}
\begin{centering}
\includegraphics[clip,width=0.45\textwidth]{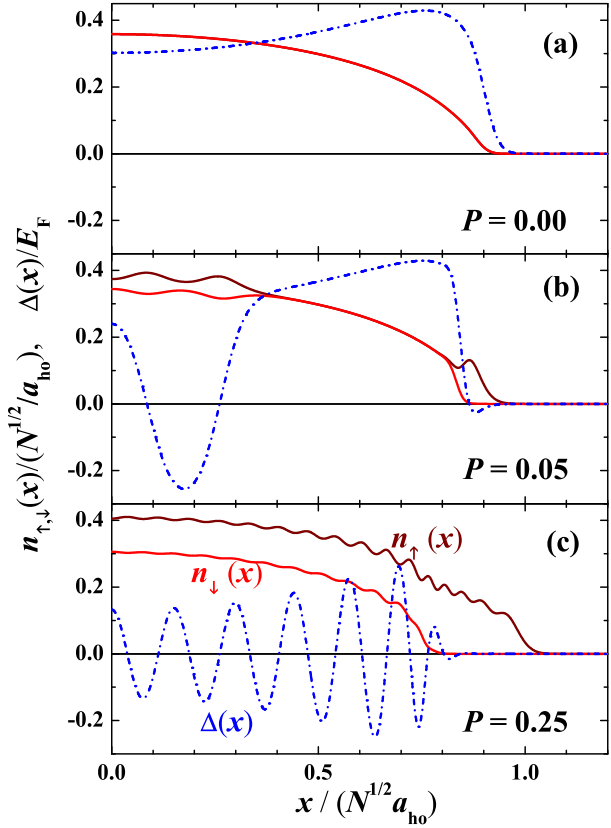}
\par\end{centering}

\caption{(Color online) Density profiles (solid lines) and order parameters
(dot-dashed lines) of a trapped Fermi gas at several total spin polarizations
as indicated. The dimensionless coupling constant at the trap center
$\gamma_{0}$ is $1.6$. With increasing the total spin polarization,
the FFLO enters gradually at center, leading to two phase separation
phases.}

\label{fig16}
\end{figure}


For a pure BCS superfluid with zero polarization (Fig. 16a), the spin
up and down density profiles coincide, and decrease monotonically
as expected. However, the order parameter is non-monotonic: it increases
slowly up to the boundary of the trap, and then drops to zero very
rapidly. A maximum at the trap edge then arises in the order parameter,
in marked contrast to the 3D cases, where the order parameter decreases
monotonically. This maximum is due to the low dimensionality of the
gas. Recall that the BCS prediction of the gap for a uniform gas $\Delta_{BCS}\simeq8\epsilon_{F}\exp[-\pi^{2}/(2\gamma)]$.
At the local position $x$, $\epsilon_{F}\propto n^{2}\left(x\right)$,
while $\gamma=2/\left[a_{1D}n(x)\right]$. As a result, the position
dependent order parameter is given by, \begin{equation}
\Delta_{BCS}\left(x\right)\propto n^{2}\left(x\right)\exp\left[-\frac{\pi^{2}}{4}a_{1D}n\left(x\right)\right],\end{equation}
 which is a product of $n^{2}\left(x\right)$ and of an exponent.
These two parts decrease and increase respectively towards the trap
edge. Particularly, the increase of the exponent is due to the increase
of the effective interactions, which becomes much larger with decreasing
density. Therefore their interplay should result in a maximum. In
general, the exponent is dominant, thereby the sharp decrease or the
maximum of $\Delta_{BCS}\left(x\right)$ occurs at the trap edge for
a moderate local density.

With increasing total spin polarization, the order parameter starts
to oscillate at the trap center, suggesting the entry of FFLO-type
states at center. Correspondingly, the spin up and down density profiles
are no longer the same. For a small total spin polarization (Fig.
16b), the oscillation of the order parameter is restricted at the
trap center, and the ordinary BCS order parameter still persists at
the edge. As a consequence, we find a phase separation phase consisting
of a FFLO state at the trap center and a standard BCS state outside.
There is also a very small region with a weak oscillation of the order
parameter, occurring exactly at the trap boundary. Presumably, it
is a finite size effect. As we shall see later, the resulting normal
cloud at the boundary is an artifact of the mean-field theory, which
turns to break down at sufficient small densities or large interactions.

Increasing further the spin polarizations (Fig. 16c), the oscillations
of the order parameter penetrate the whole cloud. We find then another
phase separation phase, with an interior core of a FFLO superfluid
phase and an outer shell of the normal component. Therefore, there
should be a critical total spin polarization, $P_{c}$, that separates
the two phase separation phases. The periodicity of the oscillations
in the FFLO phase can be estimated, and we find a reasonable agreement
with the single-plane-wave estimation for $q$ if we treat the gas
as locally homogeneous at the trap center.

The validity of the mean-field BdG calculations in the trap environment
will be commented later on, by comparing the mean-field density profiles
with that obtained from the exact Gaudin solution and the local density
approximation. The physical reason for the two phase separation phases
and the value of $P_{c}$, as well as the small oscillations in the
density profiles, will also be addressed.

%
\begin{figure}
\begin{centering}
\includegraphics[clip,width=0.45\textwidth]{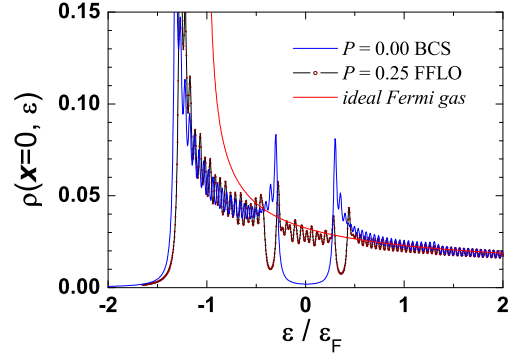}
\par\end{centering}

\caption{(Color online) Local fermionic density of states of a trapped polarized
Fermi gas at an interaction strength $\gamma_{0}=1.6$. The remarkable
two-energy-gap structure is robust in the trap environment.}

\label{fig17}
\end{figure}


Finally, we study the local fermionic density of the state in the
trap. In Fig. 17, we report the density of states at the trap center
for a BCS superfluid (a) and a FFLO superfluid (b). In the presence
of the trap, we find that the essential feature of a two-energy-gap
structure in the FFLO state is still apparent. This may provide us
a useful experimental signature to detect indirectly the FFLO states.

\section{Asymptotically exact Gaudin solutions in a harmonic trap}

For a large number of fermions, a useful method to account for an
external trapping potential traps is to use the local density approximation
\citep{hldprl1d,orso}. Together with the Gaudin solution for the
homogeneous equation of states of a polarized Fermi gas, this gives
an asymptotically exact result as long as $N\gg1$. This condition
is readily satisfied in the on-going 1D experiment, where the typical
number of atoms $N\sim100$.

The main idea of the local density approximation is that the system
can be treated locally as infinite matter with a local chemical potential.
We then partition the cloud into many cells in which the number of
fermions is much greater than unity. Provided that the variation of
the trap potential across the cell is small compared with the local
Fermi energy, the interface effects are negligible \citep{silva1,ldabec}.
Qualitatively, the interface energy should scale like $N^{-1/d}$
compared to the total energy, where $d$ is the dimensionality.

In detail, the local density approximation amounts to determining
the chemical potential $\mu_{g}=(\mu_{\uparrow g}+\mu_{\downarrow g})/2$
and the chemical potential difference $\delta\mu_{g}=(\mu_{\uparrow g}-\mu_{\downarrow g})/2$
of the inhomogeneous gas from the local equilibrium conditions, \begin{eqnarray}
\mu_{\uparrow}\left[n(x),p(x)\right]+\frac{1}{2}m\omega^{2}x^{2} & = & \mu_{\uparrow g},\\
\mu_{\downarrow}\left[n(x),p(x)\right]+\frac{1}{2}m\omega^{2}x^{2} & = & \mu_{\downarrow g},\end{eqnarray}
 and the normalization conditions, \begin{eqnarray}
N & = & \int_{-\infty}^{+\infty}n(x)dx,\\
NP & = & \int_{-\infty}^{+\infty}n(x)p(x)dx,\end{eqnarray}
 where $n(x)$ and $p(x)$ are respectively the total linear density
and the local spin polarization, and $P$ the total spin polarization.
We have used a subscript {}``$g$'' to denote the global chemical
potentials.

To solve these equations, we rewrite the chemical potentials in the
form, \begin{eqnarray}
\mu_{\uparrow}\left[n(x),p(x)\right] & = & \frac{\hbar^{2}}{2m}n^{2}(x)\bar{\mu}_{\uparrow}\left[\gamma(x),p(x)\right],\\
\mu_{\downarrow}\left[n(x),p(x)\right] & = & \frac{\hbar^{2}}{2m}n^{2}(x)\bar{\mu}_{\downarrow}\left[\gamma(x),p(x)\right],\end{eqnarray}
 where $\bar{\mu}_{\sigma}$ are the reduced chemical potentials,
depending on the dimensionless coupling constant and local spin polarization
only. Further, it is convenient to rescale the chemical potentials,
coordinate and total linear density into a dimensionless form, \textit{i.e.},
\begin{eqnarray}
\bar{\mu}_{\sigma g} & = & \frac{\mu_{\sigma g}}{\epsilon_{b}},\\
\bar{x} & = & \frac{a_{1D}x}{a_{ho}^{2}},\\
\bar{n} & = & na_{1D}\text{.}\end{eqnarray}
 Then the local equilibrium equations and the normalization equations
can be rewritten as, \begin{eqnarray}
\frac{\bar{n}^{2}(\bar{x})}{2}\bar{\mu}_{\uparrow}\left[\gamma(\bar{x}),p(\bar{x})\right]+\frac{\bar{x}^{2}}{2} & = & \bar{\mu}_{\uparrow g},\\
\frac{\bar{n}^{2}(\bar{x})}{2}\bar{\mu}_{\downarrow}\left[\gamma(\bar{x}),p(\bar{x})\right]+\frac{\bar{x}^{2}}{2} & = & \bar{\mu}_{\downarrow g},\end{eqnarray}
 and \begin{eqnarray}
\frac{1}{\pi^{2}\gamma_{0}^{2}} & = & \int_{-\infty}^{+\infty}\bar{n}(\bar{x})d\bar{x},\\
\left(\frac{1}{\pi^{2}\gamma_{0}^{2}}\right)P & = & \int_{-\infty}^{+\infty}\bar{n}(\bar{x})p(\bar{x})d\bar{x}.\end{eqnarray}
 where $\gamma(\bar{x})=2/\bar{n}(\bar{x})$. The terms on the left-side
hand of the last two equations emphasize that the properties of the
cloud rely on two dimensionless parameters, $\gamma_{0}$ and $P$.
In particular, the coupling constant in a trap is controlled by $\gamma_{0}$,
where $\gamma_{0}\ll1$ corresponds to weak coupling, while $\gamma_{0}\gg1$
corresponds to the strongly interacting regime.

The numerical procedure for the local density approximation is straightforward.
For given parameters $\gamma_{0}$ and $P$, and initial guess for
$\bar{\mu}_{\sigma g}$, we invert the dimensionless local equilibrium
equations to find $\gamma(\bar{x})$ and $p(\bar{x})$. The chemical
potentials $\bar{\mu}_{\sigma g}$ are then adjusted slightly to enforce
number conservation, giving a better estimate for the next iterative
step. The iteration is continued until the number conditions are satisfied
within a certain range.

\subsection{Density profiles: LDA vs BdG}

%
\begin{figure}
\begin{centering}
\includegraphics[clip,width=0.45\textwidth]{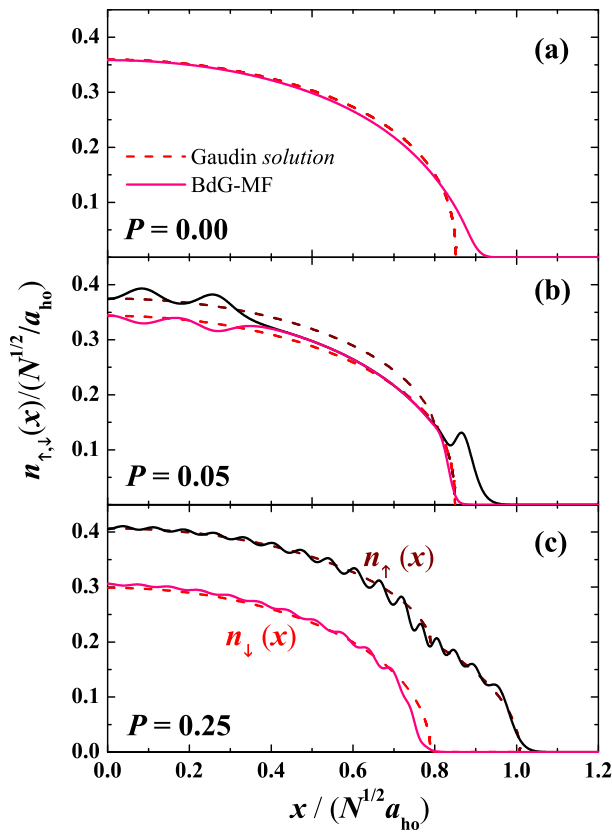}
\par\end{centering}

\caption{(Color online) Density profiles of a trapped gas, calculated by the
exact Gaudin solution and the local density approximation, are shown
for several spin polarizations as indicated. For comparison, we plot
also the self-consistent mean-field BdG predictions. They are in reasonable
agreement at the center. A discrepancy occurs at the trap edge, where
for small polarization, the approximate BdG calculation overestimates
the size of the unpolarized BCS shell.}

\label{fig18}
\end{figure}


In Fig. 18, we give the density profiles obtained from the local density
approximation using dashed lines. For comparison, we show also the
results of the BdG solutions. Apart from a negligible difference at
the trap boundary (due to a breakdown of mean-field theory), we find
a good agreement. This becomes even better as the total spin polarization
increases. In particular, the two phase separation phases found in
the BdG calculations are evident.

The appearance of the phase separation phases is easy to understand.
Within the local density approximation, the local chemical potential
$\mu(x)$ decreases parabolically away from the center of the trap
while the local chemical potential difference $\delta\mu(x)$ stays
constant. It is then evident from Fig. 15 that with a nonzero spin
polarization we always have a polarized FFLO superfluid at the trap
center where the local chemical potential (or interaction parameter)
is large (or small). Away from the center with decreasing local chemical
potential, the gas enters into either an unpolarized BCS superfluid
or a fully polarized normal cloud, depending whether the chemical
potential difference is smaller than a half of the binding energy
or not. Thus, there is a critical chemical potential difference $\delta\mu_{c}\equiv\epsilon_{b}/2$
that separates the inhomogeneous system into two phase separation
states: a mixture of a polarized superfluid core and an unpolarized
superfluid shell (FFLO-BCS), or a coexistence of a polarized superfluid
at the center and a fully polarized normal gas outside (FFLO-BCS).

It should be noted that the former phase separation phase is exotic,
as the BCS-like superfluid state occurs at the edge of the trap, in
marked contrast to the 3D case. This is caused by the peculiar effects
of low dimensionality, for which the gas becomes more nonideal with
decreasing 1D density towards the edge of the trap, and hence the
energy required to break the pairs approaches $\epsilon_{b}/2$ from
below. As $\delta\mu_{g}<\epsilon_{b}/2$, there should be a fully
paired region once the local critical chemical potential $\delta\mu_{c,p=0}>\delta\mu_{g}$,
\textit{i.e.}, the BCS-like superfluid.

Though the basic feature of the BdG results is well reproduced by
the local density approximation calculations, we note that there are
still some discrepancies that merit careful examination. First, with
decreasing the density the mean-field theory seems to fail at the
trap edge, as shown in Figs. (18a) and (18b). For a small polarization
$P=0.05$ (Fig. 18b), a notable discrepancy thus occurs at the trap
edge. The very small unpolarized BCS shell, roughly from $0.80N^{1/2}a_{ho}$
to $0.85N^{1/2}a_{ho}$ as predicted by the LDA calculation, becomes
strongly overestimated by the mean-field calculation. Secondly, there
are small oscillations in the BdG density profiles. Presumably, these
oscillations, observed also in a box with periodic boundary conditions,
are either due to the presence of the FFLO states or due to a finite
size effect. Considering the absence of the true long-range order
in 1D, we prefer the later interpretation, and regard them as the
Friedel oscillations caused by the residual unpaired atoms. To check
this point, in the BdG calculations we have varied the total number
of fermions, while keeping other parameters invariant. The oscillations
become less pronounced with increasing numbers of atoms. We emphasize
that in the on-going experiments, the total number of atoms is about
one hundred. Therefore, the oscillations in the density profiles could
be observed experimentally. However, they may hardly be considered
as a fundamental signature of the presence of the FFLO states.

\subsection{Phase diagram of a polarized Fermi gas in traps}

%
\begin{figure}
\begin{centering}
\includegraphics[clip,width=0.45\textwidth]{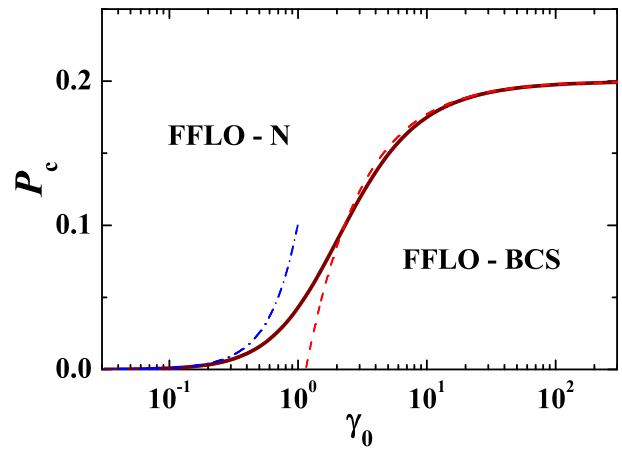}
\par\end{centering}

\caption{(Color online) Phase diagram of a one-dimensional trapped spin-polarized
Fermi gas. The dashed line and dot-dashed line are the asymptotic
results for the critical spin polarization in the strongly and weakly
interacting regimes, respectively.}

\label{fig19}
\end{figure}


We may determine numerically the critical spin polarization $P_{c}$
from the critical chemical potential difference $\delta\mu_{c}=\epsilon_{b}/2$.
In Fig. 19, we present $P_{c}$ as a function of the interaction coupling
constant $\gamma_{0}$, giving rise to a phase diagram of the inhomogeneous
polarized 1D Fermi gas \citep{hldprl1d,orso}. Again, the asymptotic
behavior of $P_{c}$ may be computed analytically in the weak and
strong coupling limits. These are shown in the figure using a dashed
line and a dot-dashed line, respectively.

Consider first a strongly interacting gas with $\gamma(x)\geq\gamma_{0}\gg1$.
Using the asymptotic expression for the chemical potential and chemical
potential difference, the rescaled local equilibrium equations can
be rewritten as, \begin{eqnarray}
-\frac{1}{2}+\frac{\pi^{2}\bar{n}^{2}\left(\bar{x}\right)}{32}\left[1-p\left(\bar{x}\right)\right]^{2}+\frac{\bar{x}^{2}}{2} & = & \bar{\mu}_{g},\nonumber \\
\frac{1}{2}+\frac{\pi^{2}\bar{n}^{2}\left(\bar{x}\right)}{32}A\left[p\left(\bar{x}\right)\right]+\frac{\pi^{2}\bar{n}^{3}\left(\bar{x}\right)}{4}B\left[p\left(\bar{x}\right)\right] & = & \delta\bar{\mu}_{g},\end{eqnarray}
 where $A[p(\bar{x})]=-1+2p\left(\bar{x}\right)+15p^{2}\left(\bar{x}\right)$
and $B\left[p\left(\bar{x}\right)\right]=-p\left(\bar{x}\right)/4+9p^{2}\left(\bar{x}\right)/2-67p^{3}\left(\bar{x}\right)/12$.
Note that in this limit $\bar{n}\left(\bar{x}\right)\ll1$. In the
rescaled units, the critical chemical potential difference $\delta\bar{\mu}_{g}$
is exactly $1/2$. Therefore, if we consider up to $A[p(\bar{x})]$
only in the expansion, we find that the local spin polarization should
satisfy, \begin{equation}
15p^{2}\left(\bar{x}\right)+2p\left(\bar{x}\right)-1=0,\end{equation}
 which yields $p\left(\bar{x}\right)\equiv1/5$ and hence the total
spin polarization $P_{c}=1/5$. The improvement to the next order
requires the inclusion of the term $B\left[p\left(\bar{x}\right)\right]$.
For this purpose, we assume $p\left(\bar{x}\right)=1/5-\delta(\bar{x})$,
where $\delta(\bar{x})\ll1$. The summation of $A\left[p\left(\bar{x}\right)\right]$
and $B\left[p\left(\bar{x}\right)\right]$ terms should be zero at
the critical polarization. Thus, to leading order of $\delta(\bar{x})$,
we find that, \begin{equation}
\delta(\bar{x})=\frac{32}{375}\bar{n}\left(\bar{x}\right).\end{equation}
 The density profile $\bar{n}\left(\bar{x}\right)$ can be determined
by using the local equilibrium equation for $\bar{\mu}_{g}$, which
to a good approximation \begin{equation}
-\frac{1}{2}+\frac{\pi^{2}}{50}\bar{n}^{2}\left(\bar{x}\right)+\frac{\bar{x}^{2}}{2}=\bar{\mu}_{g}.\end{equation}
 Combined with the normalization condition, $\int_{-\infty}^{+\infty}\bar{n}(\bar{x})d\bar{x}=1/(\pi^{2}\gamma_{0}^{2})$,
we find that, \begin{equation}
\bar{n}\left(\bar{x}\right)=\frac{\sqrt{10}}{\pi^{2}\gamma_{0}}\left[1-\frac{5\pi^{2}\gamma_{0}^{2}\bar{x}^{2}}{2}\right]^{1/2}.\end{equation}
 Therefore, we determine the critical spin polarization using $P_{c}=\pi^{2}\gamma_{0}^{2}\int_{-\infty}^{+\infty}\bar{n}\left(\bar{x}\right)p\left(\bar{x}\right)d\bar{x}$
and find that, \begin{eqnarray}
P_{c} & = & \frac{1}{5}-\frac{256}{225\pi^{3}}\sqrt{\frac{2}{5}}\frac{1}{\gamma_{0}},\nonumber \\
 & = & 0.2-\frac{0.023208}{\gamma_{0}}.\end{eqnarray}

The consideration in the weak coupling limit is much simple. In the
rescaled units, \begin{equation}
\delta\bar{\mu}\left[\bar{n}\left(\bar{x}\right),p\left(\bar{x}\right)\right]=\frac{\pi^{2}\bar{n}^{2}\left(\bar{x}\right)p\left(\bar{x}\right)}{4}=\delta\bar{\mu}_{g},\end{equation}
 where in this limit $\bar{n}\left(\bar{x}\right)\gg1$. By setting
$\delta\bar{\mu}_{g}=1/2$, we then obtain, \begin{equation}
p\left(\bar{x}\right)=\frac{2}{\pi^{2}}\frac{1}{\bar{n}^{2}\left(\bar{x}\right)}.\end{equation}
 Using again the normalization condition for the total number of atoms,
the rescaled (ideal) density profile takes the form, \begin{equation}
\bar{n}\left(\bar{x}\right)=\frac{2}{\pi^{2}\gamma_{0}}\left[1-\pi^{2}\gamma_{0}^{2}\bar{x}^{2}\right]^{1/2}.\end{equation}
 Thus, by integrating out $P_{c}=\pi^{2}\gamma_{0}^{2}\int_{-\infty}^{+\infty}2/[\pi^{2}\bar{n}\left(\bar{x}\right)]d\bar{x}$,
we find that, \begin{equation}
P_{c}=\frac{\gamma_{0}^{2}}{\pi^{2}}.\end{equation}

\section{Conclusions and some remarks}

In conclusion, we have presented a systematic study of an attractive
polarized atomic Fermi gas in one dimension, both in free space and
in a harmonic trap. The theoretical approaches include the (asymptotically)
exact Bethe ansatz solution and two mean-field approximations: the
single-plane-wave approximation for the order parameter and the self-consistent
Bogoliubov-de Gennes equations. These useful tools provide us with
quantitative phase diagrams in both uniform and harmonic trapped systems.
Our main results may be summarized as follows, in response to the
theoretical issues raised in the Introduction:

(A) We have clarified the structure of the one-dimensional FFLO states
in a uniform gas. For small spin polarization, the FFLO order parameter
behaves like a lattice of instantons and anti-instantons, which carry
the excess unpaired atoms. For a large spin polarization, the singularity
of the instantons merges together. Thus, the form of the order parameter
becomes a cosine function, as originally proposed by Larkin and Ovchinnikov
\citep{lo}. The nodes in the FFLO order parameter lead to a two-energy-gap
structure in the local fermionic density of states, which may be experimentally
observable using spectroscopic methods.

(B) We have determined the nature of the phase transition from a BCS
superfluid state to a FFLO phase. It is a smooth second order transition.
As a consequence, a one-dimensional phase separation does not occur
for a homogeneous gas. Turning to the trapped case, we find two exotic
phase separation phases. However, these phase separations are simply
trap effects.

(C) We have checked the validity of the two mean-field approaches
in the weakly or moderately interacting regimes, by comparing the
results with the exact or asymptotically exact Bethe ansatz solutions.
The mean-field methods are found to provide a useful description in
these regimes. In particular, by comparing the equations of state
and density profiles, we have shown that the spin polarized superfluid
in the Bethe ansatz solution corresponds to an FFLO state, with a
real (cosine-like) order parameter. This correspondence, however,
does not hold quantitatively in the strongly interacting regime. The
Bethe ansatz solutions do not result in any abrupt changes for the
polarized superfluid, as the interaction strengths increase from the
weak to strong regimes.

Though our study is restricted here to the one-dimensional case, we
can still obtain some insight into the phase diagram of a three-dimensional
polarized Fermi gas. This is under strong debate at the moment. Two
remarks may be in order in this respect.

One key remark is that the FFLO window in three dimension can be expected
to be much larger than that obtained from mean-field calculations
with a single-plane-wave assumption for the order parameter. As we
have noted, by improving the form of the order parameter to the Larkin
and Ovchinnikov (LO) type, $\Delta({\bf x})\propto\cos[{\bf q\cdot x}]$,
Yoshida and Yip have indeed found recently that the FFLO state becomes
more stable \citep{yip3}. Further, the one-dimensional results indicate
that one might expect a smooth phase transition from the BCS state
to FFLO state in three dimensions, although clearly this needs to
be checked with a full three-dimensional calculation.

Another interesting issue concerns the existence of a phase separation
in a three dimensional homogeneous polarized gas. From the one-dimensional
calculations, we do not find any strong indication for this. Accordingly,
the experimentally observed phase separation may simply be understood
as a trap effect. We note, however, that the three dimensional strongly
interacting BEC limit has no correspondence in the one-dimensional
attractive polarized gas \citep{tokatly,zwerger}. It that limit,
a homogeneous polarized superfluid, which may be called the Sarma
phase, becomes stable \citep{srprl,yip03,yip06}. This phase has a
different symmetry from the spatially inhomogeneous FFLO phase. Therefore,
there could be another phase intervening between the Sarma phase and
the FFLO phase. This may be a possible reason for the observation
of phase separation in three dimension. If this exists, we would expect
that phase separation for a homogeneous gas would be restricted to
the strongly-interacting regime near unitarity.


\begin{acknowledgments}
This work was supported by an Australian Research Council Center of
Excellence grant, the National Natural Science Foundation of China
Grants Nos. NSFC-10574080 and NSFC-10774190, and the National Fundamental
Research Program Grants Nos. 2006CB921404 and 2006CB921306.
\end{acknowledgments}


\begin{thebibliography}{100}
\bibitem{FR} S. Inouye \textit{et al.}, Nature (London) \textbf{392},
151 (1998).

\bibitem{lattice} M. Greiner \textit{et al.}, Nature (London) \textbf{415},
39 (2002).

\bibitem{huinp} H. Hu, P. D. Drummond, and X.-J. Liu, Nature Physics
\textbf{3}, 469 (2007), and references therein.

\bibitem{leggett} A. J. Leggett, \textit{Modern Trends in the Theory
of Condensed Matter} (Springer-Verlag, Berlin, 1980).

\bibitem{nsr} P. Nozi\`{e}res, and S. Schmitt-Rink, J. Low Temp. Phys.
\textbf{59}, 195 (1985).

\bibitem{randeria} J. R. Engelbrecht, M. Randeria, and C. A. R. Sá de
Melo, Phys. Rev. B \textbf{55}, 15153 (1997).

\bibitem{griffin} Y. Ohashi and A. Griffin, Phys. Rev. Lett. \textbf{89},
130402 (2002).

\bibitem{hui04} H. Hu \textit{et al.}, Phys. Rev. Lett. \textbf{93},
190403 (2004).

\bibitem{hld} H. Hu, X.-J. Liu, and P. D. Drummond, Europhys. Lett.
\textbf{74}, 574 (2006).

\bibitem{jila} C. A. Regal and D. S. Jin, Phys. Rev. Lett. \textbf{92},
040403 (2004).

\bibitem{mit04} M. W. Zwierlein \textit{et al.}, Phys. Rev. Lett.
\textbf{92}, 120403 (2004).

\bibitem{duke04} J. Kinast \textit{et al.}, Phys. Rev. Lett. \textbf{92},
150402 (2004).

\bibitem{ins04a} M. Bartenstein \textit{et al.}, Phys. Rev. Lett.
\textbf{92}, 203201 (2004).

\bibitem{ins04b} C. Chin \textit{et al.}, Science \textbf{305}, 1128
(2004).

\bibitem{ens} T. Bourdel \textit{et al.}, Phys. Rev. Lett. \textbf{93},
050401 (2004).

\bibitem{duke05a} J. Kinast \textit{et al.}, Science \textbf{307},
1296 (2005).

\bibitem{mit05} M. W. Zwierlein \textit{et al.}, Nature (London)
\textbf{435}, 1047 (2005).

\bibitem{duke05b} J. E. Thomas, J. Kinast, and A. Turlapov, Phys.
Rev. Lett. \textbf{95}, 120402 (2005).

\bibitem{mit06} J. K. Chin, Nature (London) \textbf{443}, 961 (2006).

\bibitem{duke07} L. Luo \textit{et al.}, Phys. Rev. Lett. \textbf{98},
080402 (2007).

\bibitem{randy} G. B. Partridge \textit{et al.}, Phys. Rev. Lett.
\textbf{95}, 020404 (2005).

\bibitem{mit06a} M. W. Zwierlein \textit{et al.}, Science \textbf{311},
492 (2006).

\bibitem{mit06b} M. W. Zwierlein \textit{et al.}, Nature (London)
\textbf{442}, 54 (2006)

\bibitem{mit06c} Y. Shin\textit{\ et al.}, Phys. Rev. Lett. \textbf{97},
030401 (2006).

\bibitem{mit07} C. H. Schunck \textit{et al.}, Science
\textbf{316}, 867 (2007)

\bibitem{rice06a} G. B. Partridge \textit{et al.}, Science \textbf{311},
503 (2006).

\bibitem{rice06b} G. B. Partridge \textit{et al.}, Phys. Rev. Lett.
\textbf{97}, 190407 (2006).

\bibitem{ff} P. Fulde and R. A. Ferrell, Phys. Rev. \textbf{135},
A550 (1964).

\bibitem{lo} A. I. Larkin and Y. N. Ovchinnikov. Zh. Eksp. Teor.
Fiz. \textbf{47}, 1136 (1964) {[}Sov. Phys. JETP \textbf{20}, 762
(1965)].

\bibitem{rmp} For a review on the FFLO states, see, for example,
R. Casalbuoni and G. Nardulli, Rev. Mod. Phys. \textbf{76}, 263 (2004).

\bibitem{cecoin5} H. A. Radovan \textit{et al.}, Nature (London)
\textbf{425}, 51 (2003); A. Bianchi \textit{et al.}, Phys. Rev. Lett.
\textbf{91}, 187004 (2006); C. Matin \textit{et al.}, Phys. Rev. B
\textbf{71}, 020503 (R) (2005).

\bibitem{sarma} G. Sarma, J. Phys. Chem. Solids \textbf{24}, 1029
(1963).

\bibitem{yip03} S.-T. Wu and S.-K. Yip, Phys. Rev. A \textbf{67},
053603 (2003).

\bibitem{yip06} C.-H. Pao, S.-T. Wu, and S.-K. Yip, Phys. Rev. B.
\textbf{71}, 132506 (2006).

\bibitem{dfs02} H. Müther and A. Sedrakian, Phys. Rev. Lett. \textbf{88},
252503 (2002).

\bibitem{dfs05} A. Sedrakian \textit{et al.}, Phys. Rev. A \textbf{72},
013613 (2005).

\bibitem{dfs06} A. Sedrakian, H. Müther and, and A. Polls, Phys.
Rev. Lett. \textbf{97}, 140404 (2006).

\bibitem{bp} W. V. Liu and F. Wilczek, Phys. Rev. Lett. \textbf{90},
047002 (2003).

\bibitem{bedaque} P. F. Bedaque, H. Caldas, and G. Rupak, Phys. Rev.
Lett. \textbf{91}, 247002 (2003).

\bibitem{son} D. T. Son and M. A. Stephanov, Phys. Rev. A \textbf{74},
013614 (2006).

\bibitem{mannarelli} M. Mannarelli, G. Nardulli, and M. Ruggieri,
Phys. Rev. A \textbf{74}, 033606 (2006).

\bibitem{yang1} K. Yang, arXiv:cond-mat/0508484.

\bibitem{yang2} K. Yang, arXiv:cond-mat/0603190.

\bibitem{srprl} D. E. Sheehy and L. Radzihovsky, Phys. Rev. Lett.
\textbf{96}, 060401 (2006).

\bibitem{sraop} D. E. Sheehy and L. Radzihovsky, Annals of Physics \textbf{322}, 1790
(2007).

\bibitem{chevy1} F. Chevy, Phys. Rev. Lett. \textbf{96}, 130401 (2006).

\bibitem{chevy2} F. Chevy, Phys. Rev. A \textbf{74}, 063628 (2006).

\bibitem{hui06} H. Hu and X.-J. Liu, Phys. Rev. A \textbf{73}, 051603(R)
(2006).

\bibitem{xiaji06} X.-J. Liu and H. Hu, Europhys. Lett. \textbf{75},
364 (2006).

\bibitem{hui07} H. Hu, X.-J. Liu, and P. D. Drummond, Phys. Rev.
Lett. \textbf{98}, 060406 (2007).

\bibitem{xiaji07} X.-J. Liu, H. Hu, and P. D. Drummond, Phys. Rev.
A \textbf{75}, 023614 (2007).

\bibitem{yip1} S.-T. Wu, C.-H. Pao, and S.-K. Yip, Phys. Rev. B \textbf{74},
224504 (2006).

\bibitem{yip2} C.-H. Pao and S.-K. Yip, J. Phys.: Condens. Matter
\textbf{18}, 5567 (2006).

\bibitem{yip3} N. Yoshida and S.-K. Yip, Phys. Rev. A \textbf{75}, 063601
(2007).

\bibitem{parish} M. M. Parish, F. M. Marchetti, A. Lamacraft, and
B. D. Simons, Nature Physics \textbf{3}, 124 (2007).

\bibitem{lobo} C. Lobo, A. Recati, S. Giorgini, and S. Stringari,
Phys. Rev. Lett. \textbf{97}, 200403 (2006).

\bibitem{torma1} J. Kinnunen, L. M. Jensen, and P. Törm\"{a}, Phys. Rev.
Lett. \textbf{96}, 110403 (2006).

\bibitem{torma2} T. Koponen, J. Kinnunen, J.-P. Martikainen, L.M.
Jensen, P. Törm\"{a}, New J. Phys. \textbf{8}, 179 (2006).

\bibitem{torma3} L. M. Jensen, J. Kinnunen, and P. Törm\"{a}, arXiv:cond-mat/0604424.

\bibitem{bulgac1} A. Bulgac, M. M. Forbes, and A. Schwenk, Phys.
Rev. Lett. \textbf{97,} 020402 (2006).

\bibitem{bulgac2} A. Bulgac and M. M. Forbes, Phys. Rev. A \textbf{75},
031605(R) (2007).

\bibitem{levin1} C.-C. Chien, Q. Chen, Y. He, and K. Levin, Phys.
Rev. Lett. \textbf{97}, 090402 (2006).

\bibitem{levin2} C.-C. Chien, Q. Chen, Y. He, and K. Levin, Phys.
Rev. A \textbf{74}, 021602(R) (2006).

\bibitem{carlson} J. Carlson and S. Reddy, Phys. Rev. Lett. \textbf{95},
060401 (2005).

\bibitem{lianyihe} L. He, M. Jin, and P. Zhang, Phys. Rev. B \textbf{73,}
214527 (2006); \textbf{74}, 024516 (2006); \textbf{74}, 214516 (2006).

\bibitem{caldas1} H. Caldas, arXiv:cond-mat/0601148.

\bibitem{caldas2} H. Caldas, arXiv:cond-mat/0605005.

\bibitem{ho} T.-L. Ho and H. Zhai, J. Low Temp. Phys. \textbf{148}, 33
(2007).

\bibitem{gu} Z.-C. Gu, G. Warner, and F. Zhou, arXiv:cond-mat/0603091.

\bibitem{iskin} M. Iskin and C. A. R. Sá de Melo, Phys. Rev. Lett.
\textbf{97}, 100404 (2006).

\bibitem{duan1} W. Yi and L.-M. Duan, Phys. Rev. A \textbf{73}, 031604(R)
(2006).

\bibitem{duan2} W. Yi and L.-M. Duan, Phys. Rev. A \textbf{73}, 063607
(2006).

\bibitem{duan3} W. Yi and L.-M. Duan, Phys. Rev. A \textbf{74}, 013610
(2006).

\bibitem{duan4} G.-D. Lin, W. Yi, and L.-M. Duan, Phys. Rev. A \textbf{74},
031604 (2006).

\bibitem{silva1} T. N. De Silva and E. J. Mueller, Phys. Rev. A \textbf{73},
051602(R) (2006).

\bibitem{silva2} T. N. De Silva and E. J. Mueller, Phys. Rev. Lett.
\textbf{97}, 070402 (2006).

\bibitem{stoof1} M. Haquea and H. T. C. Stoof, Phys. Rev. A \textbf{74},
011602(R) (2006).

\bibitem{stoof2} K. B. Gubbels, M. W. J. Romanns, and H. T. C. Stoof,
Phys. Rev. Lett. \textbf{97}, 210402 (2007).

\bibitem{stoof3} M. Haquea and H. T. C. Stoof, Phys. Rev. Lett. \textbf{98}, 260406
(2007).

\bibitem{ldabec} A. Imambekov \textit{et al.}, Phys. Rev. A \textbf{74},
053626(R) (2006).

\bibitem{martikainen} J.-P. Martikainen, Phys. Rev. A \textbf{74},
013602 (2006).

\bibitem{castorina} P. Castorina \textit{et al.}, Phys. Rev. A \textbf{72},
025601 (2005).

\bibitem{machida1} T. Mizushima, K. Machida, and M. Ichioka, Phys.
Rev. Lett. \textbf{94}, 060404 (2005).

\bibitem{machida2} K. Machida, T. Mizushima, and M. Ichioka, Phys.
Rev. Lett. \textbf{97}, 120407 (2006).

\bibitem{hldprl1d} H. Hu, X.-J. Liu, and P. D. Drummond, Phys. Rev.
Lett. \textbf{98}, 070403 (2007).

\bibitem{gaudin} M. Gaudin, Phys. Lett. \textbf{24A}, 55(1967).

\bibitem{takahashi} M. Takahashi, Prog. Theor. Phys. \textbf{44},
348(1970).

\bibitem{krivnov} V. Ya. Krivnov and A. A. Ovchinnikov, Zh. Eksp.
Teor. Fiz. \textbf{67}, 1568 (1974) {[}Sov. Phys. JETP \textbf{40},
781 (1975)].

\bibitem{guan1} M. T. Batchelor \textit{et al.,} Journal of Physics
Conference Series \textbf{42}, 5 (2006).

\bibitem{guan2} X.-W. Guan \textit{et al.}, Phys. Rev. B
\textbf{76}, 085120 (2007).

\bibitem{xiaji1d} X.-J. Liu, P. D. Drummond, and H. Hu, Phys. Rev.
Lett. \textbf{94}, 136406 (2005).

\bibitem{orso} G. Orso, Phys. Rev. Lett. \textbf{98}, 070402 (2007).

\bibitem{yang1d} K. Yang, Phys. Rev. B \textbf{63}, 140511(R) (2001),
and references therein.

\bibitem{mf1d} K. Machida and H. Nakanishi, Phys. Rev. B \textbf{30},
122 (1984).

\bibitem{buzdin} A. I. Buzdin and S. V. Polonskii, Zh. Eksp. Teor.
Fiz. \textbf{93}, 747 (1987) {[}Sov. Phys. JETP \textbf{66}, 422 (1987)].

\bibitem{esslinger1} H. Moritz, T. Stöferle, M. Köhl, and T. Esslinger,
Phys. Rev. Lett. \textbf{91}, 250402 (2003).

\bibitem{esslinger2} H. Moritz, T. Stöferle, K. Gunter, M. Köhl,
and T. Esslinger, Phys. Rev. Lett. \textbf{94}, 210401 (2005).

\bibitem{peter} K. V. Kheruntsyan \textit{et al.}, Phys. Rev. Lett.
\textbf{91}, 040403 (2003); P. D. Drummond \textit{et al.}, \textit{ibid}.
\textbf{92}, 040405 (2004).

\bibitem{karen} K. V. Kheruntsyan and P. D. Drummond, Phys. Rev.
A \textbf{61}, 063816 (2000); S. J. J. M. F. Kokkelmans \textit{et
al.}, Phys. Rev. A \textbf{65}, 053617 (2002); P. D. Drummond and
K. V. Kheruntsyan, Phys. Rev. A \textbf{70}, 033609 (2004).

\bibitem{diener} R. Diener and T.-L. Ho, arXiv:cond-mat/0405174.

\bibitem{xiaji} X.-J. Liu and H. Hu, Phys. Rev A \textbf{72}, 063613
(2005).

\bibitem{bergeman} T. Bergeman, M. G. Moore, and M. Olshanii, Phys.
Rev. Lett. \textbf{91}, 163201 (2003).

\bibitem{astrakharchik} G. E. Astrakharchik, D. Blume, S. Giorgini,
and L. P. Pitaevskii, Phys. Rev. Lett. \textbf{93}, 050402 (2004).

\bibitem{footnote} Note the difference in the definition of $a_{\rho}$
with \citep{bergeman}, which accounts for $A=-\zeta(1/2)/\sqrt{2}\simeq1.0326$.

\bibitem{tokatly} I. V. Tokatly, Phys. Rev. Lett. \textbf{93}, 090405
(2004).

\bibitem{zwerger} J. N. Fuchs, A. Recati, and W. Zwerger, Phys. Rev.
Lett. \textbf{93}, 090408 (2004).

\bibitem{lieb} E. H. Lieb and W. Liniger, Phys. Rev. \textbf{130},
1605 (1963).

\bibitem{zwerger03} W. Zwerger, J. Opt. B: Quantum Semiclass. Opt.
\textbf{5}, S9 (2003).

\bibitem{a3dB} M. Bartenstein \textit{et al.}, Phys. Rev. Lett. \textbf{94},
103201 (2005).

\bibitem{bdg} P. de Gennes, {\em Superconductivity of Metals and
Alloys} (Addison-Wesley, New York, 1966).

\bibitem{reidl} J. Reidl \textit{et al.}, Phys. Rev. A \textbf{59},
3816 (1999).
\end{thebibliography}
\end{document}